\documentclass[preprint,prd,aps,floatfix,preprintnumbers,superscriptaddress,bibnotes,nofootinbib]{revtex4-1}

\usepackage{epsfig}
\usepackage{amsmath}
\usepackage{latexsym}
\usepackage[psamsfonts]{amssymb}
\usepackage{graphicx}
\usepackage{ulem}
\usepackage{longtable}
\usepackage{epstopdf}
\usepackage{bm}% bold math
\usepackage{color}
\usepackage{multirow}
\usepackage{tablefootnote} % footnote for Table
\usepackage[colorlinks=true, linkcolor=blue, citecolor=blue, urlcolor=blue]{hyperref}
\usepackage{footmisc}
\newcommand{\be}{\begin{equation}}
\newcommand{\ee}{\end{equation}}
\newcommand{\bea}{\begin{eqnarray}}
\newcommand{\eea}{\end{eqnarray}}
\newcommand{\bi}{\begin{itemize}}
\newcommand{\ei}{\end{itemize}}

\begin{document}
%\preprint{****}
%-- title ---

\title{
Lattice study of scattering phase shifts for $DD^*$ and $BB^*$ systems using twisted boundary conditions: Search for bound state formation
}

%-- author list ---
\author{Masato Nagatsuka\:}
\email[E-mail: ]{masato.nagatsuka.r4@dc.tohoku.ac.jp}
\affiliation{Department of Physics, Tohoku University, Sendai 980-8578, Japan}
\author{Shoichi~Sasaki\:}
\email[E-mail: ]{ssasaki@nucl.phys.tohoku.ac.jp}
\affiliation{Department of Physics, Tohoku University, Sendai 980-8578, Japan}
%
%\collaboration{PACS Collaboration}

\date{\today}
%%%%%%%%%%%%% ABSTRACT %%%%%%%%%%%%%%%%%%%%%%
%-- abstract ---
\begin{abstract}
We investigate the $S$- and $P$-wave phase shifts
for the $DD^\ast$ and $BB^\ast$ scatterings using L\"uscher's finite-size method under twisted boundary conditions to search for doubly charmed tetraquaks $T_{cc}^+$ and doubly bottomed tetraquarks $T_{bb}^-$ as the hadronic bound states.
The $T_{cc}^+$ state was observed as a peak just below the $DD^*$ threshold by LHCb Collaboration, while the $T_{bb}^-$ state is a theoretically predicted tetraquark state having heavier quark flavors $bb\bar u \bar d$. L\"uscher's finite-size method is one of the well-established methods for calculating the scattering phase shifts between two hadrons in lattice QCD simulations. 
Several studies have used simulations under the periodic boundary condition to determine the scattering phase shifts at a few discrete momenta for the $DD^*$ system. However, the scattering phase shift has not been investigated for the $BB^*$ system.
In this study, we investigate $S$- and $P$-wave scattering phase shifts for the $DD^*$ and $BB^*$ systems in both $I=0$ and $I=1$ channels under several types of partially twisted boundary conditions. 
The use of the partially twisted boundary conditions enables us to obtain the scattering phase shift at any momentum by continuously varying the twisting angle. It also allows us to easily access the $P$-wave scattering phase shifts through the mixing of $S$ and $P$ waves, which is induced by the imposed boundary conditions.
The 2+1 flavor PACS-CS gauge ensembles at $m_\pi=295$, 411, and 569 MeV are used. For charm and 
bottom quarks, the relativistic heavy quark action is
adopted to reduce the lattice discretization artifacts
due to the heavy quark mass.
We discuss the emergence of a shallow bound state with a binding energy of $\mathcal{O}(100)$ keV at the physical pion mass in the $BB^*$ system, which has the quantum number $I(J^P)=0(1^+)$.
\end{abstract}

%%%%%%%%%%%%%%%%%%%%%%%%%%%%%%%%%%%%%%%%%%
\maketitle

%--- main text -------------------------------------------------------
 
%%%%%%%%%%%%%%  SEC 1  %%%%%%%%%%%%%%%%%%%%%%%%
\section{Introduction}
\label{sec:INTRO}

After the discovery of the exotic meson $X$(3872) in 2003~\cite{Belle:2003nnu}, many hidden heavy exotic hadrons were reported, and these hadrons have been studied extensively over the past several decades. However, the presence of quark-antiquark annihilation makes theoretical analyses of most exotic mesons quite challenging.
From a theoretical perspective, therefore, the existence of exotic hadrons without quark-antiquark annihilation is highly desirable. 
The experimental observation of the $S=+1$ baryon resonance, called the $\Theta^+ (1540)$ resonance~\cite{LEPS:2003wug}, was the first candidate for a genuine exotic pentaquark state classified as a $uudd\bar{s}$ state, and triggered many lattice QCD calculations~\cite{{Sasaki:2003gi},{Csikor:2003ng},{Sasaki:2004vz}}.
However, subsequent experiments have not confirmed the existence of the $\Theta^+(1540)$ state~\cite{Liu:2014yva}.

The doubly charmed tetraquark $T_{cc}^+$, which has the explicit exotic flavor $cc\bar u\bar d$ and isospin $I(J^P)=0(1^+)$, was discovered by the LHCb Collaboration in 2021~\cite{LHCb:2021vvq,LHCb:2021auc}.
The $T_{cc}^+$ state was found to be about 360 keV below the $D^0D^{*+}$ threshold with a decay width of 48 keV~\cite{LHCb:2021auc}.
The doubly bottomed tetraquark $T_{bb}^-$ can be obtained by replacing the charm quarks with bottom quarks as a theoretical analog of the $T_{cc}^+$.
The properties of $T_{cc}^+$ and $T_{bb}^-$ have been studied within the same framework
using phenomenological methods since the 1980s~\cite{Carlson:1987hh, Manohar:1992nd}
and there have been many recent updates
~\cite{Meng:2020knc, Noh:2021lqs, Dong:2021bvy, Deng:2021gnb, Kim:2022mpa, Maiani:2022qze,
Praszalowicz:2022sqx, Dai:2022ulk, Ortega:2022efc, Dai:2023kwv, Feijoo:2023sfe, Ren:2023pip,
Sakai:2023syt, Wu:2024zbx}.
The $bb\bar u\bar d$ channel is typically considered to have a single bound state with a binding energy ranging from $\mathcal{O}(1)$ to $\mathcal{O}(100)$ MeV. Furthermore, an exact few-body calculation suggests that there may be two bound states in this channel~\cite{Meng:2020knc}.

The $DD^{\ast}$ scattering has been widely studied in lattice QCD simulations to identify the $T_{cc}^{+}$ state as a shallow bound state.
The $S$-wave scattering phase shifts for the $I(J^P)=0(1^+)$ channel have been calculated using L\"uscher's finite-size method~\cite{Padmanath:2022cvl, Collins:2024sfi, Chen:2022vpo, Prelovsek:2025vbr} and the HAL QCD method~\cite{Ikeda:2013vwa,Lyu:2023xro}, while there are a few calculations for the $I(J^P)=1(1^+)$ channel~\cite{Chen:2022vpo, Meng:2024kkp}.
Although both methods are well-established approaches for obtaining the scattering phase shifts, the conclusion regarding the $T_{cc}^+$ from lattice QCD remains controversial due to the left-hand cut of $DD\pi$ just below the $DD^*$ threshold, which could give rise to the singularity in the $DD^*$ scattering amplitude as discussed in Ref.~\cite{Du:2023hlu}.
Some studies circumvent the issue by incorporating the $DD\pi$ scattering, as demonstrated in Refs.~\cite{Dawid:2024dgy, Hansen:2024ffk}.
The authors of Refs.~\cite{Collins:2024sfi, Meng:2024kkp} attempt to extract the singular behavior by assuming that the left-hand cut originates from one-pion exchange. An alternative approach involves extracting $S$- and $P$-wave scattering phase shifts based on chiral effective field theory~\cite{Meng:2023bmz}.
In this study, however, we will extract the detailed scattering phase shifts of the $S$- and $P$-wave scattering near the threshold using twisted boundary conditions, which, in principle, can provide scattering phase shifts at arbitrary energies.

There are some lattice studies for the doubly bottomed tetraquark using the nonrelativistic QCD (NRQCD) action, which show the existence of a deeply bound state with a binding energy of $\mathcal{O}(100)~\mathrm{MeV}$ in the $I(J^P)=0(1^+)$ channel~\cite{Francis:2016hui, Junnarkar:2018twb, Leskovec:2019ioa, Mohanta:2020eed, Hudspith:2023loy, Aoki:2023nzp, Alexandrou:2024iwi, Tripathy:2025vao}.
Unlike most model calculations, which predicted a wide range of binding energies from $\mathcal{O}(1)$ to $\mathcal{O}(100)$ MeV, these lattice calculations consistently predict strong binding of $B$ and $B^\ast$ mesons in the $bb\bar{u}\bar{d}$ system.
While the presence of a deep energy level from threshold is sufficient for identifying a deep bound state, the identification of shallow bound states is better served by using L\"uscher's finite-size method to probe the peculiar behavior of the scattering phase shifts in the vicinity of threshold. 
As summarized in a recent review of lattice QCD calculations for the heavy tetraquark states~\cite{Francis:2024fwf}, the identification of a deep bound state has only been discussed in the $BB^\ast$ channel. However, the search for shallow bound states, as is done in the $DD^\ast$ channel, has never been discussed.

The finite-size method is frequently utilized to examine the low-energy interaction between two hadrons.
Detailed information of the low-energy interaction is 
represented by the lower partial-wave phase shifts at low energies.
After original proposals given by L\"uscher~\cite{Luscher:1986pf, Luscher:1990ux}, many extensions have been developed in various ways. In particular, this formalism was first extended to moving frames with nonzero total momentum of scattered particles of equal mass~\cite{Rummukainen:1995vs}. A further extension is elaborated upon in Refs.~\cite{Fu:2011xz, Doring:2012eu, Leskovec:2012gb, Gockeler:2012yj} for the case of two hadrons with different masses.

In general, the quantity of $k^{2l+1}\cot \delta_l(k)$ for the 
$l$th-wave scattering phase shift $\delta_l(k)$ can be expanded in a power series of the scattering momentum squared $k^2$
in the vicinity of the threshold as 
\begin{equation}
k^{2l+1}\cot\delta_l(k)=\frac{1}{a_l}+\frac{1}{2}r_lk^2+v_lk^4 + {\cal O}(k^6).
\label{eq:ERE}
\end{equation}
which is called the effective range expansion~\cite{Newton:1982qc}.
Model-independent information of the low-energy interaction is 
encoded in a small set of scattering parameters, e.g., scattering length $a_0$ for the $S$ wave ($l=0$) or scattering volume $a_1$ for the $P$ wave ($l=1$).

According to the established principles of scattering theory, Levinson's theorem elegantly relates the $l$th-wave scattering phase shift at zero momentum $\delta_l(0)$ to the total number of bound states $n$ as follows: $\delta_l(0)=n\pi$~\cite{Newton:1982qc}.
Consequently, as indicated in Ref.~\cite{Sasaki:2006jn}, the behavior of the scattering phase shifts around zero momentum is of critical importance in the search for bound states.
Furthermore, the formation condition of the $S$-wave bound state
is expressed by the pole condition of the $S$ matrix as $k\cot\delta_0(k)=-\sqrt{-k^2}$ at $k=i\gamma$, where $\gamma$ denotes the binding momentum.~\footnote{
As pointed out in Ref.~\cite{Sasaki:2006jn}, this condition
is only fulfilled in the infinite-volume limit.}

In principle, the aforementioned model-independent information can be obtained by employing the finite-size method to calculate the scattering phase shifts at small scattering momenta.
However, when lattice QCD simulations are performed under periodic boundary conditions with a spatial extent of $L$, the finite-volume discretizes the momentum to $2\pi/L$ units, thereby constraining the accessible momentum. Therefore, under the standard periodic boundary conditions, the simulated scattering phase shift is not obtained continuously. There are very few simulated phase shifts near the threshold.

One way to circumvent this issue is to use the novel approach of twisted boundary conditions~\cite{Bedaque:2004kc}, which enables the treatment of any small momentum on the lattice through the variation of the twisting angle, continuously. As demonstrated in Ref.~\cite{Ozaki:2012ce}, this approach can be integrated with the finite-size method to yield more precise information in the low-energy region.
In this study, the same strategy is applied to $DD^*$ and $BB^*$ systems, which contain doubly heavy quarks suitable for the application of the twisted boundary condition. 
Our preliminary results were first reported in Ref.~\cite{Nagatsuka:2024vfl}.

This paper is organized as follows. 
In Sec.~\ref{sec:TheorFram}, we give a short outline for the theoretical derivation of 
L\"uscher's finite-size formula under twisted boundary conditions.
Section~\ref{sec:SimDetails} is devoted to give the details of parameters and
interpolating operators used in our lattice QCD simulations.
In Sec.~\ref{sec:NumResults}, we present the numerical results of the $S$- and $P$-wave scattering phase shifts of the $DD^*$ and $BB^*$ systems for 
both isospin $I=0$ and $I=1$ channels on three gauge ensembles with $m_\pi=295$, 411, and 569 MeV.
We then systematically examine the behavior of bound state formation in the $DD^*$ and $BB^*$ systems based on the $S$-wave phase shifts near the threshold.
Finally, we close with a summary in Sec.~\ref{sec:summary}.

%%%%%%%%%%%%%%  SEC 2  %%%%%%%%%%%%%%%%%%%%%%%%
\section{Theoretical Framework}
\label{sec:TheorFram}
\subsection{L\"uscher's finite-size formula}
We first recapitulate L\"uscher's finite-size formula to calculate the scattering phase shift between two hadrons. In Ref.~\cite{Luscher:1990ux},
L\"uscher originally
derived the formula for two identical massive spinless particles in the center-of-mass (c.m.) frame, relating the $S$-wave ($l=0$) scattering phase shift $\delta_0(k)$
to the discrete energy levels in a box of volume $L^3$ under the periodic boundary conditions in spatial directions. The projection into the trivial irreducible representation (irrep) of the cubic symmetry, 
corresponding to vanishing total momentum, yields
the following simple form of L\"uscher's finite-size formula for the $S$-wave phase shift near the threshold, where the higher partial-wave ($l\ge4$) contributions are safely ignored~\cite{Luscher:1990ux}:
%
% Eq.(1)
%
\begin{align}
\cot\delta_0(k)&=\frac{1}{\pi^{3/2}q}\mathcal{Z}_{00}(1;q^2)\qquad \mbox{with}\qquad q=\frac{Lk}{2\pi},
\label{eq:Luscher_formula_origin}
\end{align}
where $k$ denotes the scattering momentum of
the two-particle system in the c.m. frame.
Here, the function $\mathcal{Z}_{00}(1;q^2)$ is the generalized zeta function, which is formally defined with spherical harmonics $Y_{lm}(\hat{\bm n})$ ($\hat{\bm n}={\bm n}/|{\bm n}|$) as
%
% Eq.(2)
%
\begin{align}
\mathcal{Z}_{lm}(s;q^2)&=\sum_{{\bm n}\in \mathbb{Z}^3}
%\frac{\mathcal{Y}_{lm}({\bm n})}
\frac{|{\bm n}|^l {Y}_{lm}(\hat{\bm n})}{({\bm n}^2-q^2)^{s}}
\label{eq:zeta_function}
\end{align}
for $s>3/2$, and then an analytic continuation of the function is needed from the region $s>3/2$ to
$s=1$~\cite{{Luscher:1990ux},{Luscher:1986pf}}.

After the derivation 
for two identical particles with vanishing total momentum (c.m. frame), many different kinds of extensions have been made, e.g., 
identical bosons with nonzero total momentum (moving frame)~\cite{Rummukainen:1995vs} and two particles of unequal mass for moving frame including fermion-boson systems~\cite{Fu:2011xz, Doring:2012eu, Leskovec:2012gb, Gockeler:2012yj}.
Although such extensions may help to compute the scattering phase shifts at more values of lower scattering momenta within a given volume, one of the authors extended the formula  for the zero total momentum case with partially twisted boundary conditions to expand the range of accessible momenta, especially near the threshold, within a single volume~\cite{Ozaki:2012ce}.

\subsection{Finite-size formula under twisted boundary conditions}

Let us briefly review the derivation of the finite-size formula for the zero total momentum with twisted boundary conditions.
Following the original derivation in Ref.~\cite{Luscher:1990ux}, we assume that  
the two-particle interaction
has finite interaction range $R$.
In the exterior region of the interaction range ($|{\bm x}|>R$), the relative two-particle wave function $\Psi({\bm x})$ obeys the Helmholtz equation as
%
% Eq.(3)
%
\begin{equation}
(\nabla^2+k^2)\Psi({\bm x})=0,
\label{eq:helmholz_eqn}
\end{equation}
where $\Psi({\bm x})$ denotes the relative wave function and $k$ is the relative momentum.
We here consider the following twisted boundary condition 
imposed on the wave function $\Psi({\bm x})$ in a box with volume $L^3$ as
%
% Eq.(4)
%
\begin{equation}
\Psi({\bm x}+L{\bm e}_i)=e^{i\theta_i}\Psi({\bm x})\label{eq:twisted_boundary},
\end{equation}
where $\bm{e}_i$ %$\pmb{e}_i$
are unit vectors in the direction of the $i$ axis ($i=x,y,z$) and $\theta_i$ are real valued twisting angles, which are restricted to the range of $-\pi<\theta_{i}\leq \pi$. 
We also use a vector expression, namely, $\bm{\theta}=(\theta_x,\theta_y,\theta_z)$.
The momentum of the plane wave solutions is limited to $(2\pi{\bm n}+\bm{\theta})/L$ with ${\bm n}\in\mathbb{Z}^3$, if the conditions of 
Eq.~\eqref{eq:twisted_boundary} are imposed.
Therefore, the Green's function $G^{\bm{\theta}}({\bm x};k^2)$ 
under the twisted boundary conditions is given by
%
% Eq.(5)
%
\begin{align}
G^{\bm{\theta}}({\bm x};k^2)&=\sum_{{\bm r}\in \Gamma_{\bm \theta}}\frac{\exp({2\pi i{\bm r} \cdot {\bm x}/L})}{(2\pi{\bm r}/L) ^2 - k^2},\label{eq:green_function_twist}
\end{align}
where the summation is over the set $\Gamma_{\bm \theta}$, which is defined as 
%
% Eq.(6)
%
\begin{align}
\Gamma_{\bm \theta}&=\Bigl\{{\bm r}~\Bigl|
{\bm r}={\bm n}+\frac{\bm \theta}{2\pi},~{\bm n} \in \mathbb{Z}^3\Bigr\}\label{eq:gamma_theta}.
\end{align}
Using the harmonic polynomials $\mathcal{Y}_{lm}({\bm x}) =|{\bm x}|^l Y_{lm}({\hat{\bm x}})$ with $\hat{\bm x}={\bm x}/|{\bm x}|$, 
we can obtain a set of solutions 
%
% Eq.(7)
%
\begin{equation}
G_{lm}^{\bm{\theta}}({\bm x};k^2)=\mathcal{Y}_{lm}({\bm \nabla})G^{\bm{\theta}}({\bm x};k^2),
\end{equation}
which forms a complete basis suitable for expanding
$\Psi({\bm x})$ in an outer range ($|{\bm x}|>R$), where Eq.~\eqref{eq:helmholz_eqn} is satisfied, as
%
% Eq.(8)
%
\begin{equation}
\Psi({\bm x})=\sum_{l=0}^\infty\sum_{m=-l}^l v_{lm}G_{lm}^{\bm{\theta}}({\bm x};k^2)
\end{equation}
with constant coefficients $v_{lm}$. 
In an analogous way to L\"uscher's derivation~\cite{Luscher:1990ux}, $G_{lm}^{\bm{\theta}}({\bm x};k^2)$ can be expanded in spherical harmonics $Y_{lm}(\hat{\bm x})$ and spherical Bessel functions $n_l(k|{\bm x}|)$ and $j_l(k|{\bm x}|)$ as
%
% Eq.(9)
%
\begin{align}\label{eq:solution_finite}
G_{lm}^{\bm{\theta}}({\bm x};k^2)=
    \frac{(-1)^lk^{l+1}}{4\pi}\Bigl\{Y_{lm}(\hat{\bm x})n_l(k|{\bm x}|)
    +\sum_{l'=0}^\infty\sum_{m'=-l'}^{l'}\mathcal{M}_{lm,l'm'}^{\bm{\theta}}Y_{l'm'}(\hat{\bm x})j_{l'}(k|{\bm x}|)\Bigr\}.
\end{align}
We remark that the only difference from the original is the definition of the coefficients $\mathcal{M}_{lm,l'm'}^{\bm{\theta}}$, which are given by
%
% Eq.(10)
%
\begin{align}
\mathcal{M}_{lm,l'm'}^{\bm{\theta}}&=\frac{(-1)^l}{\pi^{3/2}}\sum_{j=|l-l'|}^{l+l'}\sum_{s=-j}^j\frac{i^j}{q^{j+1}}\mathcal{Z}_{js}^{\bm{\theta}}(1;q^2)^*C_{lm,js,l'm'},
\end{align}
where the coefficients $C_{lm,js,l'm'}$ are expressed in terms of Wigner's 3{\it j} symbol
%
% Eq.(11)
%
\begin{align}
C_{lm,js,l'm'}=&(-1)^{m'}i^{l-j+l'}\sqrt{(2l+1)(2j+1)(2l'+1)}\notag\\
&\times
\left(\begin{matrix}
  l & j & l' \\
  m & s & -m'  
\end{matrix}\right)
\left(\begin{matrix}
  l & j & l' \\
  0 & 0 & 0  
\end{matrix}\right)
\end{align}
and the generalized zeta function with the twisting angle vector $\bm{\theta}$ 
is defined by
%
% Eq.(12)
%
\begin{align}
\mathcal{Z}_{lm}^{\bm{\theta}}(s;q^2)&=\sum_{{\bm r}\in \Gamma_{\bm{\theta}}} 
\frac{\mathcal{Y}_{lm}({\bm r})}{({\bm r}^2-q^2)^{s}}.
\end{align}
The elements of $\Gamma_{\bm{\theta}}$ are reduced to non-negative integers, 
${\bm n}\in \mathbb{Z}^3$, when ${\bm{\theta}}={\bm0}$. Then
Eq.~\eqref{eq:solution_finite} reproduces the original equation proved in Ref.~\cite{Luscher:1990ux}.

Whereas the expression of the wave function defined in a cubic box is obtained in terms of spherical harmonics as shown in Eq.~\eqref{eq:solution_finite}, the general solution of the Helmholtz equation at $|{\bm x}|>R$ in infinite volume is known~\cite{Newton:1982qc} as
%
% Eq.(13)
%
\begin{align}\label{eq:solution_infinite}
\Psi({\bm x})=\sum_{l=0}^\infty\sum_{m=-l}^lb_{lm}Y_{lm}(\hat{\bm x})\{\alpha_l(k) j_l(k|{\bm x}|)
+\beta_l(k)n_l(k|{\bm x}|)\}
\end{align}
with constant coefficients $b_{lm}$. The coefficients $\alpha_l(k)$ and $\beta_l(k)$ are related to the $l$th-wave scattering phase shift $\delta_l(k)$ as $\beta_l(k)/\alpha_l(k)=\tan\delta_l(k)$.
Since the coefficients of the independent solutions in Eqs.~\eqref{eq:solution_finite} and~\eqref{eq:solution_infinite} can be equated, we obtain the following relations:
%
% Eq.(14) and (15)
%
\begin{align}
b_{lm}\alpha_l(k)&=\sum_{l'=0}^\infty\sum_{m'=-l'}^{l'}v_{l'm'}\frac{(-1)^{l'}}{4\pi}k^{l'+1}\mathcal{M}_{l'm',lm}^{\bm{\theta}}, \label{eq:alpha} \\
b_{lm}\beta_l(k)&=v_{lm}\frac{(-1)^{l}}{4\pi}k^{l+1}. \label{eq:beta}
\end{align}
Inserting Eq.~\eqref{eq:beta} into Eq.~\eqref{eq:alpha} yields
%
% Eq.(16)
%
\begin{align}
\sum_{l=0}^\infty\sum_{m=-l}^{l}\left(\mathcal{M}_{lm,l'm'}^{\bm{\theta}}-\delta_{ll'}\delta_{mm'}\cot\delta_l(k)\right)\beta_{l}(k)b_{lm}=0.
\end{align}
Thus, the existence of nontrivial solutions for $b_{lm}$ 
requires that the following determinant vanishes:
%
% Eq.(17)
%
\begin{align}
\det\big(\mathcal{M}_{lm,l'm'}^{\bm{\theta}}-\delta_{ll'}\delta_{mm'}\cot\delta_l(k)\big)=0.
\end{align}
The matrix elements $\mathcal{M}_{lm,l'm'}^{\bm{\theta}}$ 
can be expressed by a linear operator
$\hat{\cal M}^{\bm{\theta}}$ 
\begin{align}
\mathcal{M}_{lm,l'm'}^{\bm{\theta}}
=\langle lm|\hat{\cal M}^{\bm{\theta}}|l'm'\rangle
\end{align}
with the set of basis vectors $|lm\rangle$, which spans the irreducible representations of the continuous
rotational group.

On the lattice, the continuous rotational symmetry is broken to the discrete rotational symmetry $G$, i.e., the cubic group $O_h$ or its subgroup. Depending on the choice of the twisting angle vector ${\bm \theta}$, the symmetry group $G$ will change as shown in Table~\ref{Tab:Classification}.
Therefore, the above determinant equation can be further reduced by using the irreducible representations of the corresponding cubic group, instead of the vectors $|lm\rangle$.

Since the cubic group maintains a subgroup symmetry of the continuous rotational group, the basis vectors of an irreducible representation $\Gamma$ under the discrete rotational symmetry $G$ can be expressed as the linear combination of $|lm\rangle$,
\begin{align}
|\Gamma\alpha ln\rangle = \sum_{m=-l}^lc_{lm}^{\Gamma\alpha n}|lm\rangle,
\end{align}
where $\alpha$ runs from 1 to the dimension of $\Gamma$, and
$n$ runs from 1 to the multiplicity of the irreducible representation $\Gamma$.
The coefficients of the unitary transformation $c_{lm}^{\Gamma\alpha n}$ are listed in Ref.~\cite{Gockeler:2012yj}.
The matrix elements of the operator $\hat{\cal M}^{\bm{\theta}}$ in the
new basis are given by
\begin{align}
\langle\Gamma\alpha ln|\hat{\cal M}^{\bm{\theta}}|\Gamma'\alpha' l'n'\rangle=\sum_{m=-l}^l\sum_{m'=-l'}^{l'}
c_{lm}^{\Gamma\alpha n*}c_{l'm'}^{\Gamma'\alpha' n'}\mathcal{M}^{\bm{\theta}}_{lm,l'm'}.
\end{align}
which can be partially diagonalized according to Schur's lemma~\cite{Luscher:1990ux} as
%
% Eq.
%
\begin{align}
\langle\Gamma\alpha ln|\hat{\cal M}^{\bm{\theta}}|\Gamma'\alpha' l'n'\rangle=\delta_{\Gamma\Gamma'}\delta_{\alpha\alpha'}\mathcal{M}^{\bm{\theta},\Gamma}_{ln,l'n'},
\end{align}

In this study, we focus on the $S$-wave ($l=0$) and $P$-wave ($l=1$) phase shifts for the scattering of two hadrons under the twisted boundary condition with nontrivial twisting angle vectors. 
When the scattering is considered near the threshold, the higher partial-wave contributions ($l>1$) can be safely ignored. Therefore, we set $\delta_l=0$ for $l>1$, which is justified at small $k^2$. Furthermore, the twisted boundary condition enables us to accommodate the finite relative momentum, even for the lowest Fourier mode, 
${\bm n}=(0, 0, 0)$. This may allow us to use wall-type quark fields, which are automatically projected onto the trivial 
irrep, $A_1$($A_{1g})$~\cite{{Ozaki:2012ce},{Gockeler:2012yj}}.

As pointed out in Ref.~\cite{Ozaki:2012ce}, the generalized 
Rummukainen-Gottlieb formula for the scattering of two particles with different mass
with nonzero total momentum (moving frame) can be easily translated into the L\"uscher finite-size formula given under the twisted boundary conditions in the c.m. frame, where the Lorentz boost factor is unity.
Following the original expression for a
moving frame, when we take an assumption that $\delta_l=0$ for $l>1$, the finite-size formulas of the $A_1$ irrep are given as the following determinant equation
\begin{equation}
\left|\begin{matrix}
  \cot\delta_0(k)-\mathcal{M}_{{\rm SS}}^{\bm{\theta}}(q) & \mathcal{M}_{{\rm SP}}^{\bm{\theta}}(q) \\
  \mathcal{M}_{{\rm SP}}^{\bm{\theta}}(q)^* & \cot\delta_1(k)-\mathcal{M}_{{\rm PP}}^{\bm{\theta}}(q)  
\end{matrix}\right|=0,
\label{eq:Luscher_formula_twist}
\end{equation}
which properly takes into account the mixing of
$S$- and $P$-wave phase shifts in one relation.
For the three types of the twisting angle vectors,
$\bm{\theta}=(0,~0,~\theta)$,
$(\theta,~\theta,~0)$, and $(\theta,~\theta,~\theta)$,
the explicit expression of the matrix elements $\mathcal{M}_{{\rm SS}}^{\bm{\theta}}$, $\mathcal{M}_{{\rm SP}}^{\bm{\theta}}$, and $\mathcal{M}_{{\rm PP}}^{\bm{\theta}}$
are summarized in Table~\ref{Tab:MatrixElem},
where we use the shorthand notation
\begin{align}
w_{lm}(q)&=
\frac{1}{\pi^{3/2}\sqrt{2l+1}q^{l+1}}
\mathcal{Z}_{lm}^{\bm{\theta}}(1;q^2),
\label{eq:omega_definition}
\end{align}
which implies that $w_{11}$ and $w_{22}$ are complex functions, while $w_{00}$ and $w_{22}$ are all real functions.
If special properties of ${\rm Re}(w_{11})={\rm Im}(w_{11})$ and
${\rm Re}(w_{22})=0$ are taken into account, $w_{11}$ and $w_{22}$ can be rewritten by a single real function as
$w_{11}=(1+i){\rm Re}(w_{11})$ and $w_{22}=i{\rm Im}(w_{22})$. 

For the trivial case of $\theta=0$, all three types of the determinant equation reduce properly to the original L\"usher finite-size formula~\eqref{eq:Luscher_formula_origin} due to the group theoretical constraint of $\mathcal{M}_{{\rm SP}}^{\bm{\theta}}=0$. 
In addition, when $\theta=\pi$, parity symmetry is restored and the $S$- and $P$-wave phase shifts are disentangled due to the fact that $\mathcal{M}_{{\rm SP}}^{\bm{\theta}}=0$.
In this case, Eq.~\eqref{eq:Luscher_formula_twist} gives
a finite-size formula similar to the original Rummukainen-Gottlieb formula~\cite{Rummukainen:1995vs}, but without the Lorentz factor ($\gamma=1$).

In this study, we will evaluate both $\delta_0$ and
$\delta_1$ from the determinant equation~\eqref{eq:Luscher_formula_twist}
using two types of twisting angle vectors, 
$\bm{\theta}=(0,~0,~\theta)$ and $(\theta,~\theta,~\theta)$, which provide two independent equations.

\begin{table*}[htb]
\begin{ruledtabular}
\begin{tabular}{cccccccc}
\hline
$\bm{\theta}$ & $(0,0,0)$ & $(0,0,\theta)$ & $(\theta,\theta,0)$ & $(\theta,\theta,\theta)$ & $(0,0,\pi)$ & $(\pi,\pi,0)$
&$(\pi,\pi,\pi)$ \\
Symmetry $G$ &  $O_h$  &  $C_{4v}$ & $C_{2v}$ & $C_{3v}$
& $D_{4h}$ & $D_{2h}$  & $D_{3d}$ 
\\ \hline
$\Gamma^{(l=0)}$ & $A_{1g}$& $A_{1}$ & $A_1$ & $A_1$
& $A_{1g}$ & $A_g$ & $A_{1g}$  \\ 
$\Gamma^{(l=1)}$ & $T_{1u}$ & $A_{1}\oplus E$ & $A_{1}\oplus B_1 \oplus B_2$ & $A_{1}\oplus E$
& $A_{2u}\oplus E_u$ & $B_{1u}\oplus B_{2u}\oplus B_{3u}$ & $A_{2u} \oplus E_{u}$\\ 
%$\Gamma^{(l=2)}$ & $E_g \oplus T_{2g}$ & $A_{1}\oplus  B_{1} \oplus B_{2}\oplus E $ 
%& $2A_{1} \oplus A_{2} \oplus  B_{1} \oplus B_{2}$& $A_{1}\oplus 2E$ \\
\hline
\end{tabular}
\caption{
Irreducible decomposition of the angular momentum representations $\Gamma^{(l=0)}$ and $\Gamma^{(l=1)}$ for each point group associated with the choice of twisting angle.
For the $C_{4v}$, $C_{2v}$, and $C_{3v}$ groups, both the $l=0$ ($S$-wave) and $l=1$ ($P$-wave) representations include the trivial irrep $A_1$ due to the absence of an inversion center, though the $D_{4h}$, $D_{2h}$, and $D_{3d}$
groups have parity symmetry so that the trivial irrep, $A_{1g}(A_g)$, does not contain the $l=1$ ($P$-wave) contributions. 
$g$ (gerade) or $u$ (ungerade) denotes the existence of an inversion center and indicates even or odd parity, respectively.  
}
\label{Tab:Classification}
\end{ruledtabular}
\end{table*}
\begin{table}[htb]
%\begin{table*}
\begin{center}
\begin{ruledtabular}
\begin{tabular}{cccccccc}
\hline
Twisting angle & $(0,0,\theta)$ & $(\theta,\theta,0)$ & $(\theta,\theta,\theta)$ \\
Symmetry $G$ & $C_{4v}$ & $C_{2v}$ & $C_{3v}$ \\
%label & $[001]$ & $[110]$ & $[111]$ \\
\hline
$\mathcal{M}_{{\rm SS}}^{\bm{\theta}}(q)$ & $w_{00}$ & $w_{00}$ & $w_{00}$ \\
$\mathcal{M}_{{\rm SP}}^{\bm{\theta}}(q)$ & $i\sqrt{3}w_{10}$ & $i\sqrt{6}w_{11}$ & $i3w_{10}$ \\
$\mathcal{M}_{{\rm PP}}^{\bm{\theta}}(q)$ & $w_{00}+2w_{20}$ & $w_{00}-w_{20}-i\sqrt{6}w_{22}$ & $w_{00}-i2\sqrt{6}w_{22}$ \\
\hline
\end{tabular}
\caption{Definitions of $\mathcal{M}_{{\rm SS}}^{\bm{\theta}}(q),~\mathcal{M}_{{\rm SP}}^{\bm{\theta}}(q)$, and $\mathcal{M}_{{\rm PP}}^{\bm{\theta}}(q)$ for each twisting angle ($0< |\theta| < \pi$). For simplicity, an irrelevant phase factor
is omitted in the definition of $\mathcal{M}_{{\rm SP}}^{\bm{\theta}}(q)$. 
}
\label{Tab:MatrixElem}
\end{ruledtabular}
\end{center}
%\end{table*}
\end{table}
%

%\clearpage
%%%%%%%%%%%%%%  SEC 3  %%%%%%%%%%%%%%%%%%%%%%%%
\section{Simulation Details}
\label{sec:SimDetails}
\subsection{Lattice setup}

We will investigate the bound state formation in the $DD^*$ and $BB^*$ systems by exploring scattering phase shifts at low energies.
For this purpose, we perform lattice QCD simulations using 2+1 flavor PACS-CS gauge configurations, which are generated with the Wilson-clover quark action and the Iwasaki gauge action on a lattice size of $L^3\times T = 32^3\times 64$ at a single lattice spacing.
Simulation parameters of PACS-CS gauge configurations are summarized in Table~\ref{Tab:LatParam}.
We use three ensembles of 2+1 flavor PACS-CS gauge configurations, where
the simulated pion masses correspond to $m_\pi=295,~411$, and 569 MeV as summarized in Table~\ref{Tab:Param}.

\begin{table}[htb]
%\begin{table*}
\begin{center}
\begin{ruledtabular}
\begin{tabular}{cccccccc}
\hline
$\beta$ \ & \ $a$ (fm) \ &  $L^{3} \times T$ & \ $\sim La$ (fm) & $c_{\rm SW}$ \\
\hline
$1.9$  \ & $0.0907(13)$ \ & $32^{3} \times 64$ \ & $2.9$ \ &  $1.715$  \\
\hline
\end{tabular}
\caption{Simulation parameters of $2+1$ flavor PACS-CS gauge configurations, generated using the Wilson-clover quark action and the Iwasaki gauge action: gauge coupling ($\beta=6/g^2$), lattice spacing ($a$), lattice size ($L^3\times T$), spatial extent ($La$) and improvement coefficient ($c_{\rm SW}$).
See Ref.~\cite{Aoki:2008sm} for further details.
}
\label{Tab:LatParam}
\end{ruledtabular}
\end{center}
%\end{table*}
\end{table}
\begin{table}[htb]
%\begin{table*}
\begin{center}
\begin{ruledtabular}
\begin{tabular}{cccccc}
\hline
Ensemble&  $(\kappa_{ud}, \kappa_{s})$ & $m_{\pi}$ (MeV) & No. of configs.
\\
\hline
A & (0.13770, 0.13640) & 295(2) & 799 \\
B & (0.13754, 0.13640) & 411(1) & 450 \\
C & (0.13727, 0.13640) & 569(2) & 400 \\
\hline
\end{tabular}
\caption{Summary for three ensembles of $2+1$ flavor PACS-CS gauge configurations~\cite{Aoki:2008sm} used in this study.
The masses of the pion, as listed in the table, are evaluated in this study using the Dirichlet boundary condition imposed in the time direction. 
Our results are consistent with those in Ref.~\cite{Aoki:2008sm}.
}
\label{Tab:Param}
\end{ruledtabular}
\end{center}
%\end{table*}
\end{table}

The nonperturbatively ${\cal O}(a)$ improved Wilson-clover fermions are used for up and down quarks, while the relativistic heavy quark (RHQ) action is adopted for charm and bottom quarks. The RHQ action, which is a variant of the Fermilab approach~\cite{El-Khadra:1996wdx}, is the anisotropic version of the ${\cal O}(a)$ improved Wilson-clover action with five parameters, $\kappa$, $\nu$, $r_s$, $c_B$, and $c_E$, called RHQ parameters (for more details, see Refs.~\cite{Aoki:2001ra,Kayaba:2006cg}).
The explicit form of the RHQ action is given by
%
% Eq.
%
\begin{align}
S_\mathrm{RHQ}=&\sum_{x,y}\bar{q}(x)D(x,y)q(y), \\
D(x,y)=& \delta_{x,y}-\kappa\sum_{i=1,2,3}\left\{(r_s-\nu\gamma_k)U_{x,i}\delta_{x+\hat{k},y}+(r_s+\nu\gamma_k)U_{x,i}^\dagger
\right\} \cr
&-\kappa\left\{
(r_t-\gamma_4)U_{x,4}\delta_{x+\hat{4},y}
+(r_t+\gamma_4)U^\dagger_{x,4}\delta_{x,y+\hat{4}} 
\right\}\cr
&-\delta_{x,y}c_B\kappa\sum_{i<j}\sigma_{ij}F_{ij}(x)
-\delta_{x,y}c_E\kappa\sum_{i}\sigma_{4i}F_{4i}(x)
\end{align}
with $r_t=1$.
The field strength $F_{\mu\nu}$ is defined as
%
% Eq.
%
\begin{align}
F_{\mu\nu}(x)&=\frac{1}{2}\left(C_{\mu\nu}(x)-C_{\mu\nu}^\dagger(x)\right)
\end{align}
with the clover-leaf operator given by
%
% Eq.
%
\begin{align}
C_{\mu\nu}(x)&=\frac{1}{4}\left[
U_{x,\mu}U_{x+\hat{\mu},\nu}U^\dagger_{x+\hat{\nu},\mu}U^\dagger_{x,\nu}
+U^\dagger_{x-\nu,\nu}U_{x-\hat{\nu},\mu}U_{x+\hat{\mu}-\hat{\nu},\nu}U^\dagger_{x,\mu}
\right.\cr
&\left.
+U^\dagger_{x-\hat{\mu},\mu}U^\dagger_{x-\hat{\mu}-\hat{\nu},\nu}
U_{x-\hat{\mu}-\hat{\nu},\mu}U_{x-\hat{\nu},\nu}
+U_{x,\nu}U^\dagger_{x–\hat{\mu}+\hat{\nu},\mu}U^\dagger_{x-\hat{\mu},\nu}U_{x–\hat{\mu},\mu}
\right].
%_{\rm{AH}}
\end{align}
In the specific case of light quarks, under the condition that $\nu=r_s=1$ and $c_B=c_E=c_{\rm SW}$, the RHQ action is reduced to the Wilson-clover action.

The RHQ parameters for the charm quark have been properly determined in Ref.~\cite{Kawanai:2011jt}, where the 2+1 flavor PACS-CS configurations with $m_\pi=156$ MeV are used. 
Additionally, the RHQ parameters for the bottom quark are calibrated using the same ensemble employed in Ref.~\cite{Kawanai:2011jt} at the simulated pion mass that is closest to the physical point, $m_\pi=156$ MeV. 
The parameters $r_s$, $c_B$ and $c_E$ are determined by tadpole improved one-loop perturbation theory, as demonstrated in Ref.~\cite{Kayaba:2006cg}
with a reference of the clover coefficient $c_{\rm SW}=1.715$ for light quarks. 
As for $\nu$, we use a value determined nonperturbatively, which is adjusted to reproduce the effective speed of light as unity in the dispersion relation for the spin-averaged $1S$-bottomonium state. The value of $\kappa_b$ is chosen to reproduce the experimental spin-averaged mass of $1S$-bottomonium states $M_{\rm ave}^{\rm exp}=9.4450(6)$ GeV. This was done similarly for the charm quark~\cite{Kawanai:2011jt}.
The parameters of the RHQ action, which are used for all three ensembles 
in this work, are summarized in Table~\ref{Tab:QuarkParam} for both charm and bottom quarks.

\begin{table}[htb]
%\begin{table*}
\begin{center}
\begin{ruledtabular}
\begin{tabular}{ccccccccc}
\hline
Flavor & $\kappa_{h}$ & $\nu$ & $r_{s}$ & $c_{B}$ & $c_{E}$ \\ \hline
Charm   & $0.10819$ \ & $1.2153$ \ & $1.2131$ \ & $2.0268$ \ & $1.7911$  \\
Bottom  & $0.03989$ \ & $2.9570$ \ & $2.5807$ \ & $4.0559$ \ & $2.8357$  \\
\hline
\end{tabular}
\caption{
Parameters of the RHQ action for charm and bottom quarks used in this work.
The parameter set of the charm quark was determined in Ref.~\cite{Kawanai:2011jt}.}
\label{Tab:QuarkParam}
\end{ruledtabular}
\end{center}
%\end{table*}
\end{table}

\subsection{
Implementation of twisted boundary condition for heavy quarks
}

In this study, the twisted boundary condition is applied only to the heavy  quarks ( i.e., charm and bottom quarks), while the light quarks ( i.e., up, down, and strange quarks) are subject to the periodic boundary condition. Since we use 2+1 flavor gauge configurations, the charm and bottom quarks are treated within the quench approximation, and the twist boundary condition for
the heavy quarks can be easily introduced as follows.

Suppose heavy quark fields $Q_{\bm{\theta}}({\bm x},t)$ obey the twisted boundary
condition as
\begin{equation}
Q_{\bm{\theta}} ({\bm x}+L {{\bm e}_i},t)=e^{i{\theta_i}}Q_{\bm{\theta}}({\bm x},t)
\end{equation}
for each $i$ direction. We can introduce new quark fields, $Q'({\bm x},t)$ as
\begin{equation}
Q'({\bm x},t)=e^{-i\bm{\theta} \cdot{\bm x}/L} Q_{\bm{\theta}}({\bm x},t),
\end{equation}
which satisfies the periodic condition, $Q'({\bm x}+L {\bm e}_i,t)=Q'({\bm x},t)$. This redefinition of the quark fields affects only the hopping terms that appear
in lattice fermion actions. Thus, for the Wilson-type fermions including the RHQ quarks, replacing the link variable in the hopping term as $U_{x,\mu}\to e^{i\theta_\mu a/L}U_{x,\mu}$ with $\theta_\mu=(\bm{\theta},0)$ yields a quark propagator 
with twisted boundary conditions imposed in the spatial directions.

\subsection{
Heavy-light meson interpolating operators
}

To determine the scattering phase shifts for the $DD^\ast$ and $BB^\ast$ systems, 
we use two types of the heavy-light ($Q\bar{q}$) meson interpolating operator
composed of a heavy quark ($Q=\{c, b\}$) and a light antiquark ($q=\{u, d\}$) to construct their two-point (single hadron) and four-point (two hadrons) correlation functions. One is a local bilinear operator as
%
% Eq.
%
\begin{equation}
\label{eq:meson_local}
O^{L, q}_{\alpha(Q,\Gamma)}({\bm x},t)=\bar q({\bm x},t) \Gamma Q^\prime({\bm x},t),
\end{equation}
where the Dirac matrix $\Gamma$ is chosen as $\Gamma=\gamma_5$ for the heavy-light pseudoscalar (spin-0) meson, $H=\{D,B\}$, and
$\Gamma=\gamma_i$ ($i=1, 2, 3$) for the heavy-light vector (spin-1) mesons, $H^\ast=\{D^\ast, B^\ast\}$.
For the local bilinear operator~\eqref{eq:meson_local},
a simple summation over the spatial coordinate yields
\begin{align}
\frac{1}{L^3}\sum_{\bm x}O^{L, q}_{\alpha(Q,\Gamma)}({\bm x},t)=\frac{1}{L^3}\sum_{\bm x}\bar q({\bm x},t) \Gamma Q_{\bm \theta}({\bm x},t)e^{-i{\bm \theta}\cdot{\bm x}/L}
\rightarrow O^{L, q}_{\alpha(Q,\Gamma)}({\bm p},t),
\end{align}
which can be interpreted as the Fourier transformation to a momentum representation
of the heavy-light meson interpolating operator with the finite momentum, ${\bm p}={\bm \theta}/L$, due to the twisting angle for the heavy quark. These types of interpolating operators are reserved for use with sink operators. 

An alternative interpolating operator can be defined by replacing the quark fields with the ``wall-type'' quark fields as
%
% Eq.
%
\begin{equation}
\label{eq:meson_wall}
O^{W, q}_{\alpha(Q,\Gamma)}({\bm p},t)= \frac{1}{L^6}\sum_{{\bm x}}\bar q({\bm x},t) \Gamma \sum_{{\bm y}}Q^\prime({\bm y},t),
\end{equation}
which are useful for constructing the two-hadron interpolating operator
in the trivial irreducible representations of the point groups as discussed in Ref.~\cite{Ozaki:2012ce}. In this context, these types of interpolating operators are conveniently used for source operators. Note that the wall-type quark fields for the light quarks are automatically projected onto zero momentum. On the other hand, 
the wall-type quark fields for the heavy quarks can carry finite momentum ${\bm p}={\bm \theta}/L$, since they are subject to the twisted boundary condition with ${\bm \theta}$ as
\begin{equation}
\frac{1}{L^3}\sum_{\bm y}Q'({\bm y},t)=\frac{1}{L^3}\sum_{{\bm y}}Q_{\bm \theta}({\bm y},t)e^{-i{\bm \theta}\cdot{\bm y}/L}\rightarrow Q^\prime({\bm p},t),
\end{equation}
which can be interpreted as the Fourier transformation on the level of the quark fields. 
Therefore, the wall-source operators given by Eq.~\eqref{eq:meson_wall} are also regarded as the heavy-light meson interpolating operator projected onto the finite momentum ${\bm p}$. Recall that, since the wall-source operators lose gauge covariance, 
the Coulomb gauge fixing needs to be performed at each time slice, where they are used. 

For the sake of simplicity, the following labels are used hereafter:
$\alpha(Q,\Gamma)=H$ for the $D$ ($B$) meson and 
$\alpha(Q,\Gamma)=H^\ast(i)$ for the $D^\ast$ ($B^\ast$) meson.

\subsection{
Two-hadron interpolating operators for the $DD^\ast$ and $BB^\ast$ systems
}

Next, we consider the two-hadron interpolating operator for the $HH^\ast$ system. The two-hadron interpolating operator can be constructed by a simple product of the $H$ and $H^\ast$ meson operators. 
In order to study the c.m. system, the charm quark fields in the $H$ and $H^\ast$ meson operators are assigned opposite twisting angles so that ${\bm \theta}_H=-{\bm \theta}_{H^\ast}$, thereby ensuring that the total momentum of the $HH^\ast$ system is zero. A proper projection of the isospin $I$ is necessary in constructing the two-hadron interpolating operator, since there are two different isospin states in the $HH^\ast$ system: the $I=0$ state and the $I=1$ state. Conversely, the $HH^\ast$ system does not require any spin projection, as it possesses a single total spin state with $J=1$.

First of all, the source operator of the two-hadron state in the isospin $I$ channel is defined by using the wall-source operators~\eqref{eq:meson_wall}, which are located at $t=t_{\rm src}$ as
%
% Eq.
%
\begin{align}
\label{eq:twohadron_wall}
\Omega^{W(I)}_{HH^*(i)}({\bm p},t_{\mathrm{src}})
=\frac{1}{\sqrt2}\Bigl\{O^{W,u}_{H}({\bm p},t_{\mathrm{src}})O^{W,d}_{H^*(i)}(-{\bm p},t_{\mathrm{src}})
-(-1)^I O^{W,d}_{H}({\bm p},t_{\mathrm{src}})O^{W,u}_{H^*(i)}(-{\bm p},t_{\mathrm{src}})\Bigr\},
\end{align}
where the total momentum of the two hadrons is zero if the charm quark fields in the $H$ and $H^\ast$ meson operators are assigned opposite twisting angles as ${\bm \theta}_H=-{\bm \theta}_{H^\ast}$.
However, finite ${\bm p}={\bm \theta}_H/L$ is still responsible for the nonzero relative momentum between the $H$ and $H^\ast$ mesons. 
The two-hadron operators~\eqref{eq:twohadron_wall} constructed by the wall-source
single-hadron operators are automatically projected onto the trivial irreducible representations of any point group in the two-hadron system as discussed in Ref.~\cite{Ozaki:2012ce}.
Therefore, the resulting two-hadron correlation function constructed
with the wall-source two-hadron operators~\eqref{eq:twohadron_wall} is useful in this study because we simply use the finite-size formula of the $A_1$ sector given in Eq.~\eqref{eq:Luscher_formula_twist}.

On the other hand, for the sink operator, we use the following two-hadron operators:
\begin{align}\label{eq:twohadron_local}
\Omega^{(I)}_{HH^{*}(i)}({\bm p},t)
&=\frac{1}{L^6}\sum_{{\bm x},{\bm y}}\frac{1}{\sqrt2}
\Bigl\{O^{L,u}_{H}({\bm x},t)O^{L,d}_{H^*(i)}({\bm y},t)
-(-1)^IO^{L,d}_{H}({\bm x},t)O^{L,u}_{H^*(i)}({\bm y},t)\Bigr\},\cr
\end{align}
which are constructed by a simple product of two types of the local bilinear operator~\eqref{eq:meson_local} and summed over spatial coordinates independently. The two-hadron operator~\eqref{eq:twohadron_local} is also projected onto zero total momentum, while the nonzero relative momentum ${\bm p}$ is certainly injected between the $H$ and $H^\ast$ mesons due to an assignment of opposite twisting angles on the charm quark fields in the $H$ and $H^\ast$ meson operators.

%%%%%%%%%%%%%%  SEC 4  %%%%%%%%%%%%%%%%%%%%%%%%
\section{Numerical Results}
\label{sec:NumResults}

\subsection{
Heavy-light meson spectroscopy from two-point correlation functions}

We first evaluate the single-hadron energies of the $H$ and $H^\ast$ mesons
from the two-point correlation functions composed of two types of the single-hadron operators 
defined in Sec.~\ref{sec:TheorFram} as
%
% Eq.
%
\begin{align}
\label{eq:H_correlation}
G_H({\bm p};t,t_{\mathrm{src}})&=\langle O^{L}_H ({\bm p},t) O^{W}_H (-{\bm p},t_{\mathrm{src}})^\dagger\rangle, \\
\label{eq:H_ast_correlation}
G_{H^\ast}({\bm p};t,t_{\mathrm{src}})&=\frac{1}{3}\sum_{k=1}^3\langle O^{L}_{H^\ast(k)}({\bm p},t) O^{W}_{H^\ast(k)} (-{\bm p},t_{\mathrm{src}})^\dagger\rangle.
\end{align}
where the wall-source operators~\eqref{eq:meson_wall} are used at the source time location, $t=t_{\rm src}$, while the local operators~\eqref{eq:meson_local} are used at the sink.
For the $H^\ast$ meson, we take an average over the spatial Lorentz indices $i$, which are related to the polarization direction of the spin-1 state.

It is noteworthy that the Dirichlet boundary condition is imposed in the time direction for all quarks at $t/a=0$ and 63 to
eliminate unwanted contributions across time boundaries.
The employment of this boundary condition effectively avoids the undesirable wraparound effects, which are especially problematic in systems with more than two hadrons. We compute the quark propagators with the wall sources located at different time locations $t_{\rm src}/a=6, 57$ for the case of the ensemble A to increase statistics, while only a single time location $t_{\rm src}/a=6$ is used for the case of ensembles B and C. For ensemble A, two sets of two- and four-point correlation functions are folded together to create the single correlation function. 

%
% FIG. 1
%
%\begin{figure*}[h]
\begin{figure}[htb]
 \includegraphics[width=.98\linewidth,bb=0 0 792 612,clip]{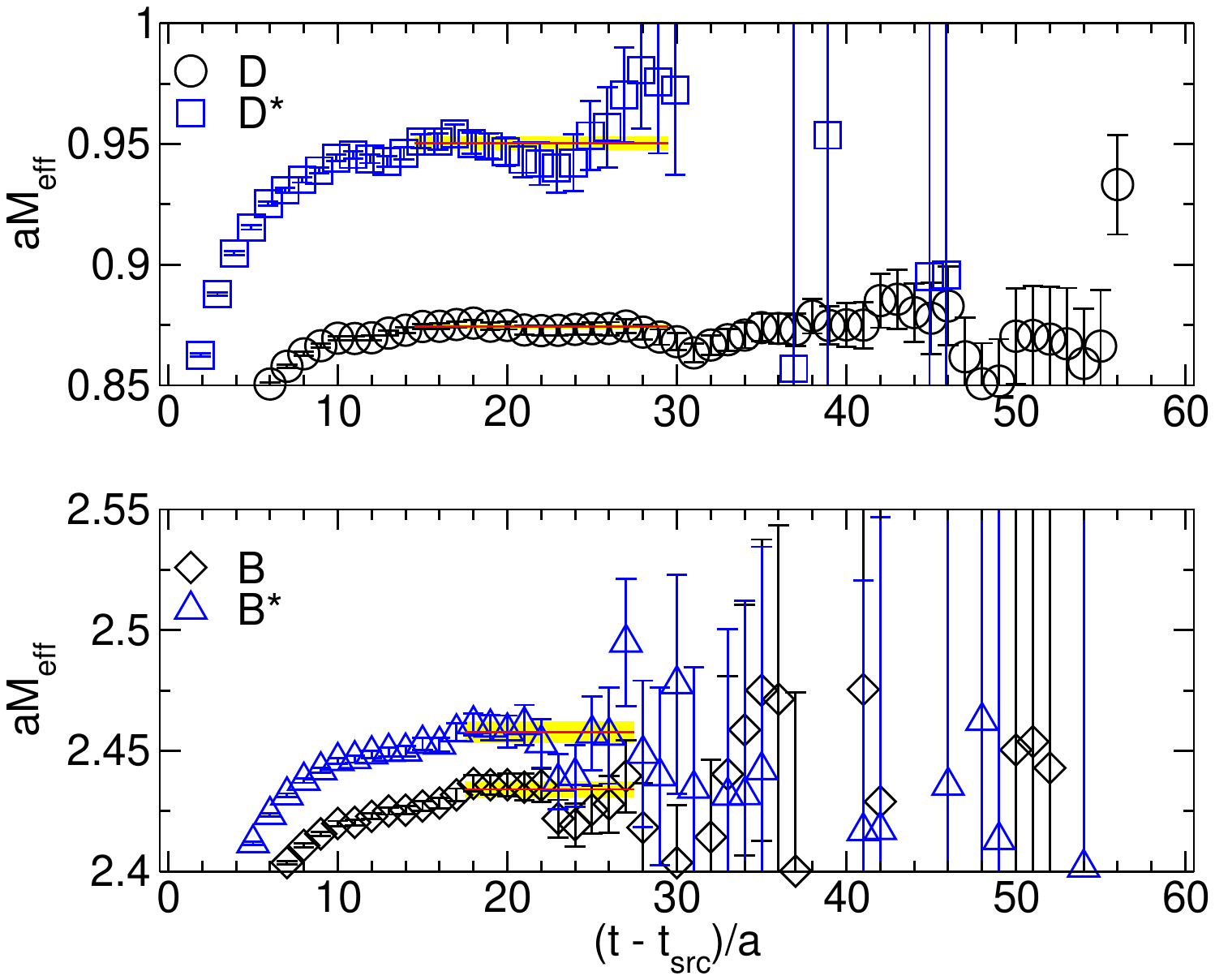}
 \caption{Effective mass plots of the $D$ and $D^*$ states (upper), and the $B$ and $B^*$ states (lower).
 Red lines indicate fit results and yellow bands display the fit ranges and
 1 standard deviation estimated by the jackknife analysis.~\label{fig:effective_mass}}
\end{figure}
%\end{figure*}
%

We can read off the single-hadron energies, $E_{X}({\bm p})$ ($X=H, H^\ast$) with the finite momentum
${\bm p}$, from the large-$t$ behavior of the two-point correlation functions under the twisted boundary condition
with the twisting angle $\bm \theta=L{\bm p}$, 
%
% Eq.
%
\begin{align}
G_{X}({\bm p};t,t_{\mathrm{src}}) \xrightarrow[t\gg t_{\mathrm{src}}]{}e^{-E_{X}({\bm p})(t-t_{\mathrm{src}})}.
\end{align}
For the ordinary periodic boundary condition (${\bm \theta}={\bm 0}$), $E_X$ is reduced to the rest mass $M_X$.
The rest masses $M_X$ of all four mesons, measured for each ensemble, are tabulated in Table~\ref{Tab:Spect}.
Figure~\ref{fig:effective_mass} shows the effective mass plots of the $D$, $D^*$, $B$, and $B^*$ mesons for ensemble A as typical examples.

\begin{figure}[htb]
 \includegraphics[width=.98\linewidth,bb=0 0 792 612,clip]{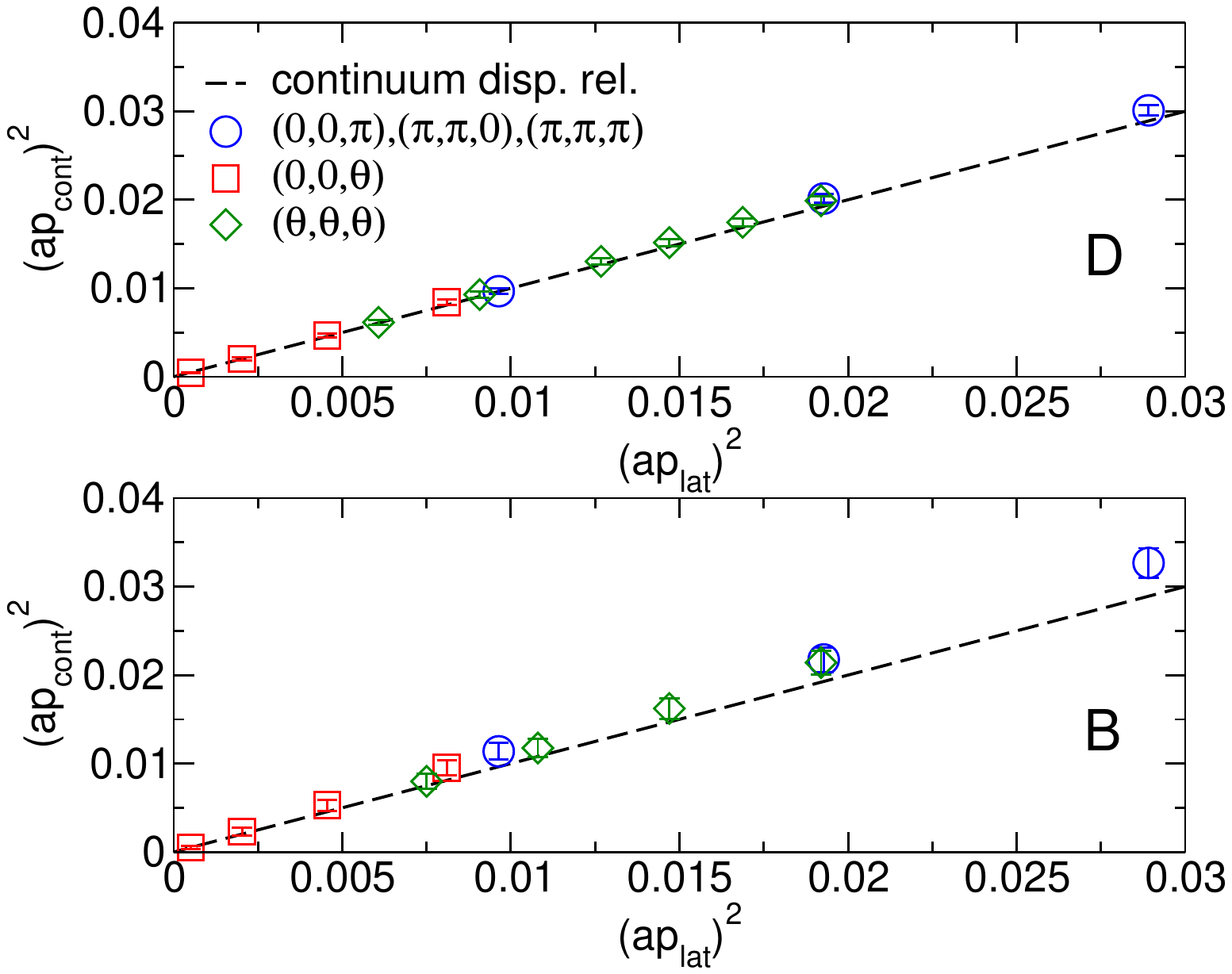}
 \caption{Check of the dispersion relations for the $D$ (upper) and $B$ (lower) states, which are calculated for ensemble A. For comparison, the continuum dispersion relation is denoted as the dotted line in each panel.~\label{fig:Disp}}
\end{figure}

Here we assume that each state on the lattice has the relativistic 
continuum dispersion relation, $E_{X}({\bm p})=\sqrt{M_X^2+{\bm p}^2}$, which is indeed enough to describe our data for heavy-light mesons with both charm and bottom quarks. In Fig.~\ref{fig:Disp}, the vertical axis shows the momentum squared defined through the relativistic continuum dispersion relation as ${\bm p}^2_{\rm con}=E_X^2-M_X^2$ for $X=D$ (upper panel) and $B$ (lower panel).
On the other hand, the horizontal axis is the momentum
squared defined by the given twisting angle as ${\bm p}_{\rm lat}^2={\bm \theta}^2/L^2$. For comparison, the relativistic continuum dispersion relation is denoted as the dotted line in each panel.
This figure shows that the large discretization errors induced by the large masses of the charm and bottom quarks are well controlled by our chosen RHQ parameters listed in Table~\ref{Tab:QuarkParam}.
Therefore, we will adopt the relativistic continuum dispersion relation to determine the scattering momentum from the energies of two-hadron states, as will be described later.

\subsection{Determination of the scattering momentum for the $HH^\ast$ system}
\label{sec:detemine_scatt_mom}

\begin{table}[htb]
%\begin{table*}
\begin{center}
\begin{ruledtabular}
\begin{tabular}{cccccc}
\hline
Ensemble & $M_{D}$ (GeV) & $M_{D^*}$ (GeV) & $M_{B}$ (GeV) & $M_{B^*}$ (GeV)
\\
\hline
A & 1.882(2) & 2.039(6) & 5.273(5) & 5.316(7) \\
B & 1.902(2) & 2.068(6) & 5.296(8) & 5.348(10) \\
C & 1.945(2) & 2.107(4) & 5.323(3) & 5.383(4)  \\
\hline
\end{tabular}
\caption{Results of the rest masses of the $D$, $D^*$, $B$, and $B^*$ mesons.}
\label{Tab:Spect}
\end{ruledtabular}
\end{center}
%\end{table*}
\end{table}
%

%
% FIG. 2
%
%\begin{figure*}[h]
%\end{figure*}
%

The scattering momentum $k_I$ of the two-hadron $HH^\ast$ system 
in the isospin $I$ channel is defined through
%
% Eq.
%
\begin{align}
W_I=\sqrt{k_I^2+M_H^2}+\sqrt{k_I^2+M_{H^*}^2},
\end{align}
where $W_I$ denotes the total energy of the two-hadron $HH^\ast$ system for isospin $I$ in the c.m. frame. The c.m. energy $W_I$ can be determined from the large-$t$ behavior of the four-point correlation functions for a two-hadron
system defined as
\begin{align}
\label{eq:two_hadron_correlation}
G^{(I)}_{HH^*}({\bm p};t,t_{\mathrm{src}})&=\frac{1}{3}\sum_{i=1}^3\langle \Omega^{(I)}_{HH^*(i)}({\bm p},t)\Omega^{W(I)}_{HH^*(i)}(-{\bm p},t_{\mathrm{src}})^\dagger\rangle \cr
&\xrightarrow[t\gg t_{\mathrm{src}}]{}e^{-W_{I}(t-t_{\mathrm{src}})}.
\end{align}

For the sake of later discussion, let us define the two-hadron energy for the isospin $I$ measured from the $HH^\ast$ threshold as 
$E_I=W_I-M$, where $M$ represents the sum of the rest masses, $M=M_H+M_{H^*}$. Then, the scattering momentum $k_I$ can be expressed in terms of the energy level $E_I$, 
\begin{align}
k_I&=\frac{\sqrt{E_I(E_I+2M)(E_I+2M_H)(E_I+2M_{H^\ast})}}{2(E_I+M)}.
\end{align}
Therefore, accurately determining $E_I$ is important for the L\"uscher finite-size analysis, for which the value of $k_I$ is
required. Instead of determining $E_I$ directly, we first calculate the energy shift $\delta E_I$ from the ratio of the two-hadron correlation function~\eqref{eq:two_hadron_correlation} and the two single-hadron correlation functions~\eqref{eq:H_correlation} and~\eqref{eq:H_ast_correlation}, 
\begin{align}
\label{eq:ratio_correlator}
R^{(I)}_{HH^*}({\bm p};t)&=\frac{G^{(I)}_{HH^*}({\bm p};t,t_{\mathrm{src}})}{G_{H}({\bm p};t,t_{\mathrm{src}})G_{H^*}({\bm p};t,t_{\mathrm{src}})}\cr
&\xrightarrow[t\gg t_{\mathrm{src}}]{}e^{-\delta E_{I}(t-t_{\mathrm{src}})},
\end{align}
which is expected to decay exponentially with the energy given by $\delta E_I = W_I-(E_H+E_{H^*})$ at large time ($t \gg t_{\rm src}$).
The ratio~\eqref{eq:ratio_correlator} can significantly reduce the statistical fluctuation due to a strong correlation between the denominator
and numerator. We then obtain the energy level $E_I$ from the following relation:
\begin{equation}
E_I=\delta E_I + (E_H - M_H) + (E_{H^*} - M_{H^*}),
\end{equation}
where the value of $E_X-M_X$ for $X=H, H^\ast$ can be evaluated only using the individual rest masses and the relativistic continuum dispersion relation. The procedure for determining $E_I$ is similar to the one proposed in Ref.~\cite{Kim:2010sd}.

\begin{figure}[htb]
 \includegraphics[width=.98\linewidth,bb=0 0 792 612,clip]{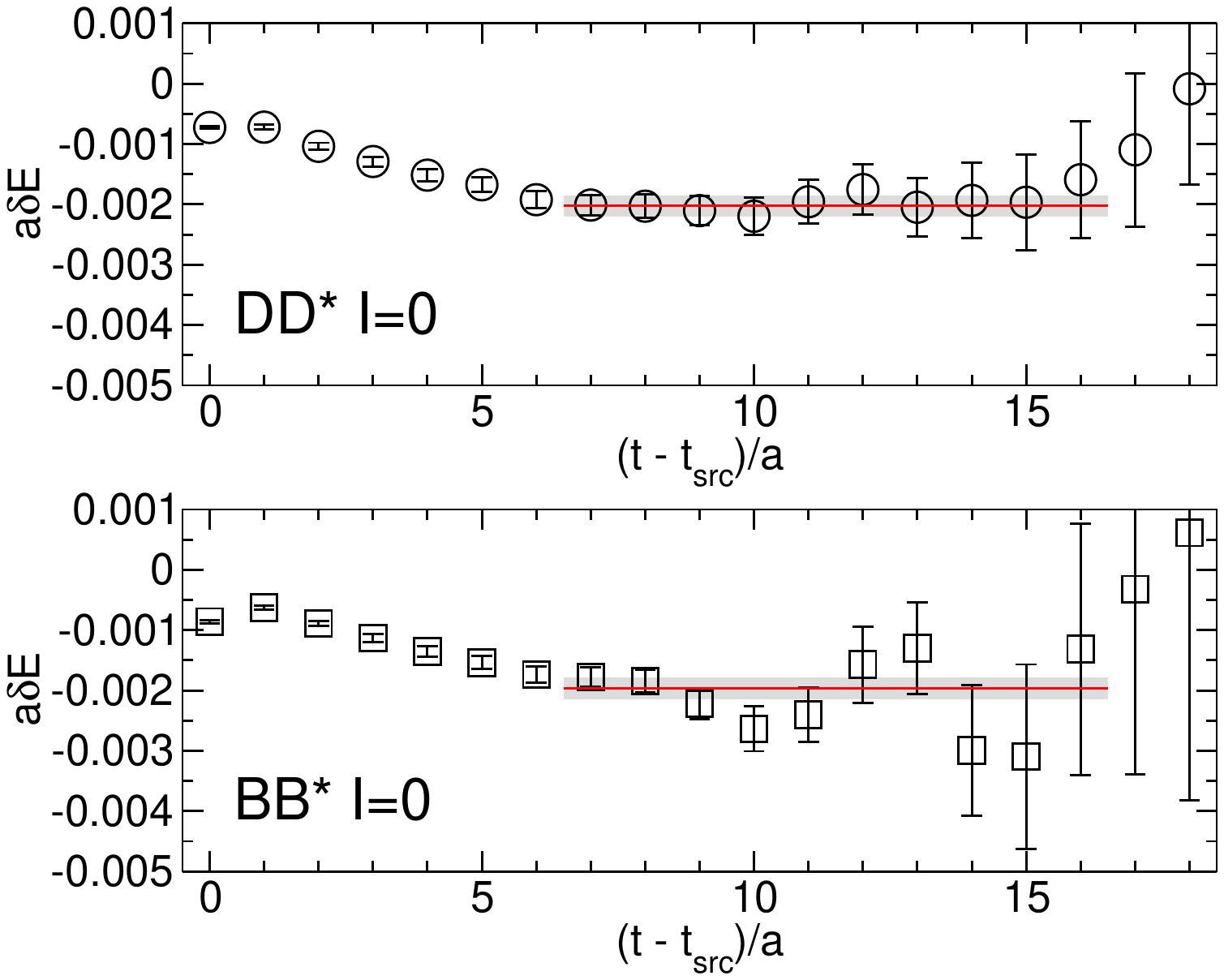}
 \caption{The effective energy shift $\delta E_{\rm eff}$ as a function of $t$ 
 for the $DD^{*}$ (upper) and $BB^{*}$ (lower) systems in the $I=0$ channel at $m_\pi=411$ MeV (ensemble B) with the trivial twisting angle, ${\bm \theta}=(0,0,0)$, corresponding the standard periodic boundary condition.
~\label{fig:effective_energy_shift}}
\end{figure}

The effective energy shift $\delta E^I_{\mathrm{eff}}$ is defined as
\begin{equation}
\delta E^I_{\mathrm{eff}}(t)=\ln\frac{R^{(I)}_{HH^*}({\bm p};t)}{R^{(I)}_{HH^*}({\bm p};t+1)},
\end{equation}
which is supposed to show a plateau for large time ($t \gg t_{\rm src}$). Figure~\ref{fig:effective_energy_shift} shows the effective energy plot of $R^{(I)}_{HH^*}({\bm p};t)$ for the $I=0$ channel of both the $DD^*$ and $BB^*$ systems. 
As a typical example, the twisting angle is set to ${\bm \theta}=(0,0,0)$ and the results are obtained using ensemble B.
In Fig.~\ref{fig:effective_energy_shift}, negative effective energy shifts are observed for both the $DD^\ast$ and $BB^\ast$ systems.
A reasonable plateau in each case is formed in the range $7 \le (t-t_{\rm src})/a \le 16$, where the two-point function of the individual mesons is also mostly dominated by the ground state of each hadron. The horizontal solid lines in Fig.~\ref{fig:effective_energy_shift} represent the fit results and the shaded bands display the fitting ranges and one standard deviation estimated by the jackknife method. 

The observed negative energy shifts for the $DD^{\ast}$ and $BB^{\ast}$ systems indicate that their lowest energies are below the threshold, though not deep enough to be interpreted as bound states. Furthermore, we calculate the energy shifts $\delta E$ for several twisting angles and then observe that $\delta E$ measured in both the $DD^\ast$ and $BB^\ast$ systems is almost unchanged with the variation of the c.m. energy $W$. 
This specific behavior is an indication of the scattering state.
Thus, the observed state is consistent with the $HH^\ast$ scattering state by a weakly attractive interaction. On the other hand, the energy shifts are found to be positive for $I=1$ in both the $DD^\ast$ and $BB^\ast$ cases. 
Therefore, as summarized in the recent review of Ref.~\cite{Francis:2024fwf}, our calculations confirm that the interaction between the $H$ and $H^*$ mesons is attractive for $I=0$ but repulsive for $I=1$.

In the next section, we will calculate the $S$- and $P$-wave scattering phase shifts, $\delta_0(k)$ and $\delta_1(k)$, respectively, for the $I=0$ and $I=1$ cases,  using the formula derived in Sec.~\ref{sec:TheorFram} with the accurate value of $k_I$ calculated from $\delta E_I$ for various twisting angles.

%
% FIG. 3
%
%\begin{figure*}[h]
%\end{figure*}
%

\subsection{Strategy for calculating $S$- and $P$-wave scattering phase shifts}
\label{sec:Strategy}

As described in the previous subsection, the energy levels, denoted by $E_I$, are obtained for each twisting angle ${\bm \theta}$ and values of the scattering momentum $k_I$ are determined.
Consequently, the scattering phase shift of both the $S$ and $P$ waves in the $HH^\ast$ system can be obtained by following the strategy outlined below.
The finite-size formula with twisted boundary conditions as expressed in Eq.~\eqref{eq:Luscher_formula_twist} contains 
the $S$- and $P$-wave phase shifts. However, for four types of special twisting angles, namely, ${\bm \theta} =(0,0,0)$, $(0,0,\pi)$, $(\pi,\pi,0)$, and $(\pi,\pi,\pi)$, Eq.~\eqref{eq:Luscher_formula_twist} is reduced to Eq.~\eqref{eq:Luscher_formula_origin}. 

This is because the even-$l$ and odd-$l$ partial waves do not mix due to the inversion symmetry of the corresponding point group, which is preserved at these special twisting angles.
As outlined in Ref.~\cite{Ozaki:2012ce},
the calculation strategy is divided into three steps, as follows:
\begin{enumerate}
%%%%% Step 1 %%%%%
\item 
First, the $S$-wave scattering phase shift $\delta_0(k)$ can be calculated independently using Eq.~\eqref{eq:Luscher_formula_origin} at the scattering momentum $k$ calculated by the twisting angles of ${\bm \theta} =(0,0,0)$, $(0,0,\pi)$, $(\pi,\pi,0)$, and $(\pi,\pi,\pi)$.
One can obtain four pairs of data ($k^2$, $\delta_0(k)$) that can be used to interpolate between the four data points via the effective range expansion formula up to ${\cal O}(k^6)$ using the jackknife method.
This treatment is justified near the threshold, where the higher partial-wave contributions ($l\ge 2$) are also safely ignored.

%%%%% Step 2 %%%%%
\item 
Next, the $P$-wave scattering phase shift $\delta_1(k)$ is evaluated at the scattering momentum $k$ calculated by the twisting angles of ${\bm \theta}=(\theta,\theta,\theta)$. The value of $\cot \delta_1(k)$ is determined by the following finite-size formula, which is derived from Eq.~\eqref{eq:Luscher_formula_twist} for the case of ${\bm \theta}=(\theta,\theta,\theta)$:  
\begin{equation}
\cot\delta_1(k)=w_{00}(q)+2\sqrt{6}{\rm{Im}}\{w_{22}(q)\}+\frac{9w_{10}^2(q)}{\cot\delta_0(k)-w_{00}(q)},
\label{eq:Luscher_formula_twist3d}
\end{equation}
where $\cot\delta_0(k)$ on the right-hand side of Eq.~\eqref{eq:Luscher_formula_twist3d} can be fixed by the interpolation data given in the first step, while 
the scaled momentum $q=\frac{kL}{2\pi}$ is determined from the energy level $E$.
Several pairs of 
$(k^2,\delta_1(k))$ are required to fit all data points.
The fitting by the effective range expansion formula up to ${\cal O}(k^4)$
provides the $P$-wave information near the threshold. 

%%%%% Step 3 %%%%%
\item 
The $S$-wave scattering phase shift $\delta_0(k)$ very close to the threshold is finally addressed by using the twisting angles of ${\bm \theta}=(0,0,\theta)$, which yields the following finite-size formula:
\begin{equation}
\cot\delta_0(k)=w_{00}(q)+\frac{3w_{10}^2(q)}{\cot\delta_1(k)-w_{00}(q)-2w_{22}(q)},
\label{eq:Luscher_formula_twist1d}
\end{equation}
where $\cot \delta_1(k)$ on the right-hand side is fixed by
the fit result of $\cot \delta_1(k)$ given by the second step.
It should be emphasized that the value of $\cot\delta_0(k)$ can be obtained at any $k^2$ in the range of $0\le k^2\le (2\pi/L)^2$ by continuously varying the twisting angle $\theta$, while it is 
only accessible at a limited number of momenta in the first step.
\end{enumerate}

This strategy has two advantages. First, the $S$-wave scattering phase shifts near the threshold can be obtained with high resolution.
Second, the $P$-wave scattering phase shift can be obtained simultaneously with the $S$-wave scattering phase shift only by calculating the irreducible representation $A_1$ of the two-hadron state.

\begin{figure*}[htb]
\centering
\includegraphics*[width=.48\textwidth,bb=0 0 792 612,clip]{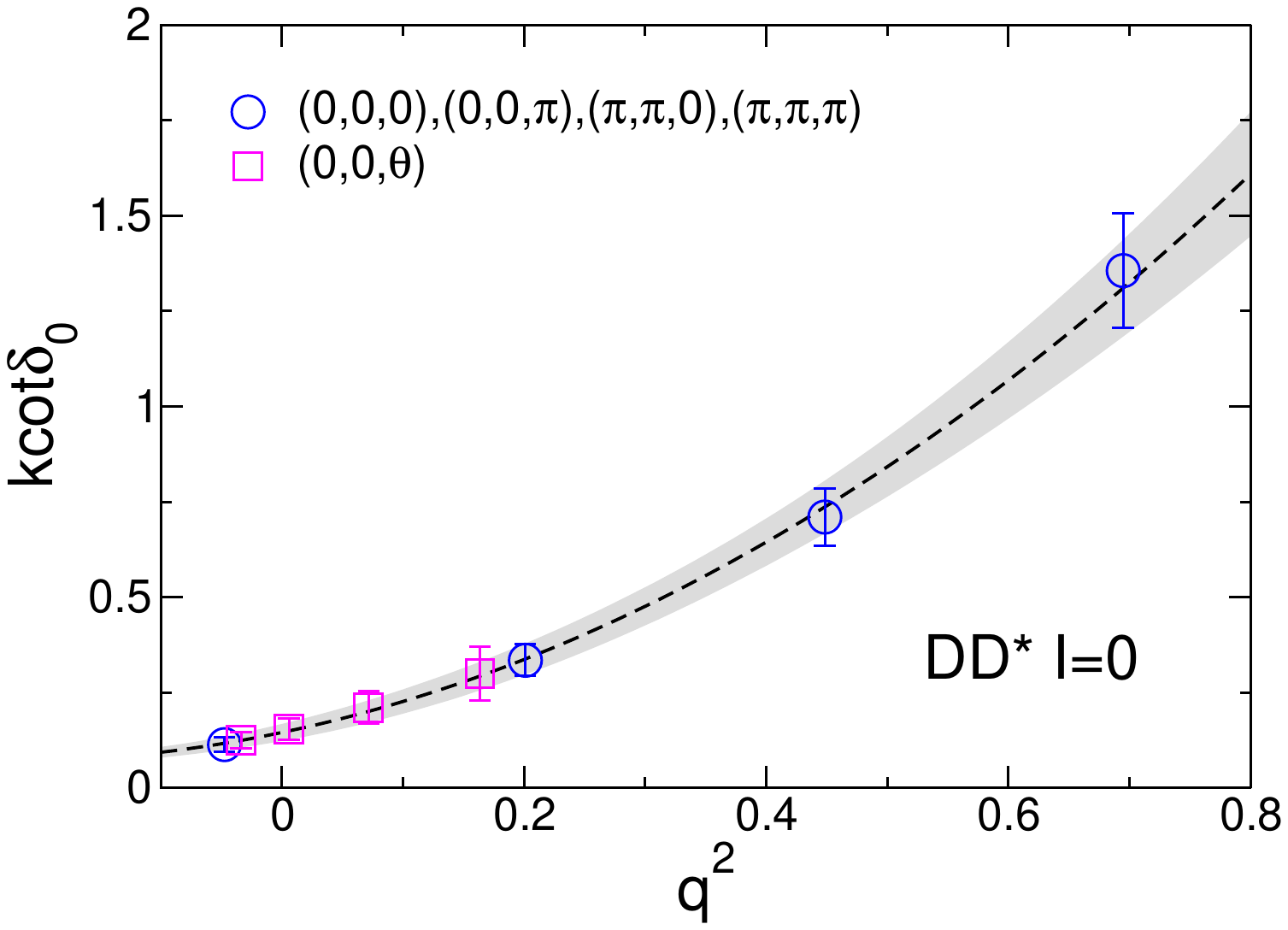} 
\includegraphics*[width=.48\textwidth,bb=0 0 792 612,clip]{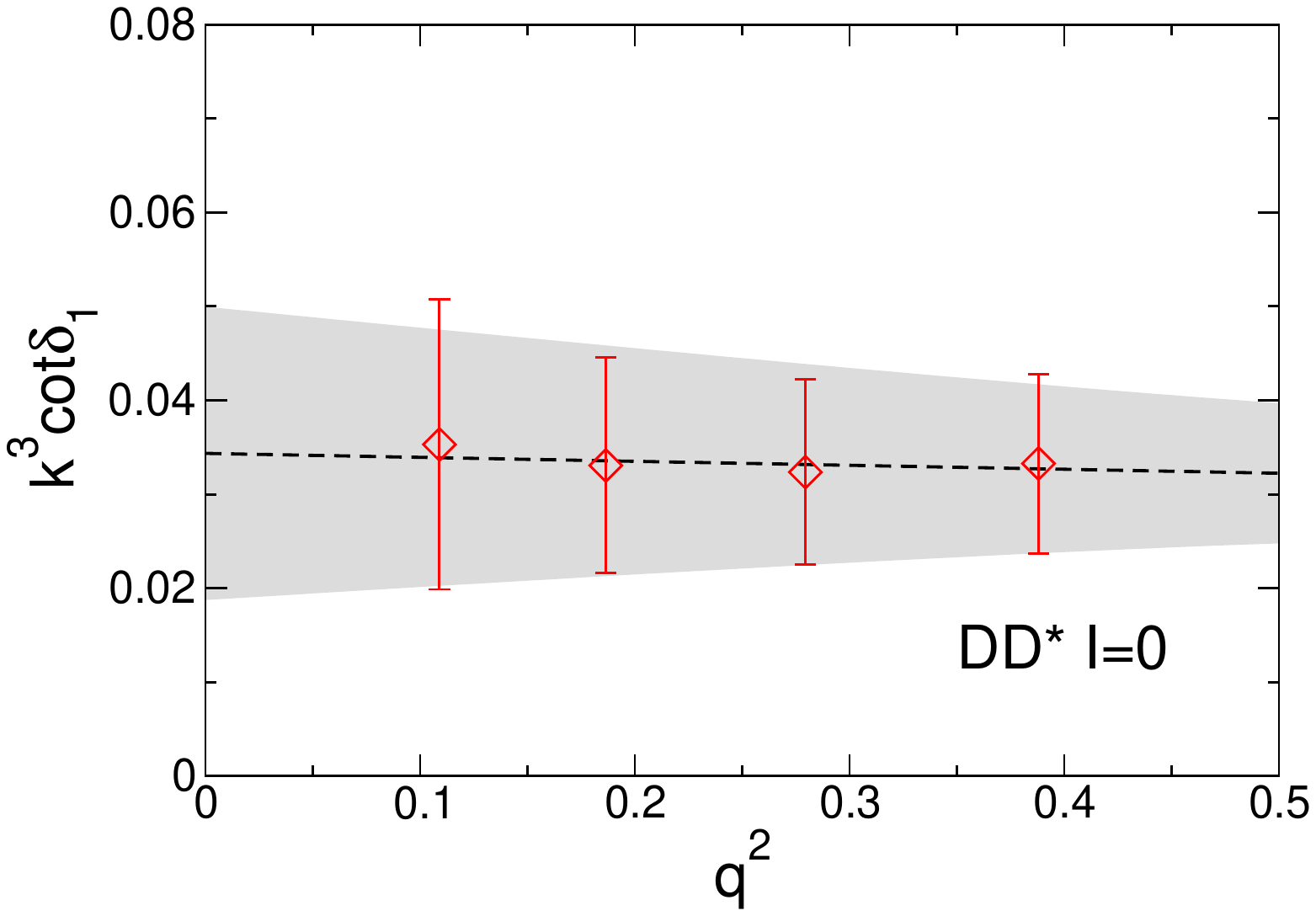}
\caption{
{\bf Left}: $k\cot\delta_0$ as a function of $q^2$ for the case of the $I=0$ $DD^*$ scattering at $m_\pi=411~\mathrm{MeV}$ (Ensemble B). 
{\bf Right}: $k^3\cot\delta_1$ as a function of $q^2$ for the case of the $I=0$ $DD^*$ scattering at $m_\pi=411~\mathrm{MeV}$ (Ensemble B).
} 
\label{fig:plot_kcot}
\end{figure*}

As a typical example, Fig.~\ref{fig:plot_kcot} displays $k\cot\delta_0$ (left panel) and $k^3\cot\delta_1$ (right panel) as functions of $q^2$ for the case of the $I=0$ $DD^*$ scattering at $m_\pi=411~\mathrm{MeV}$. The following discussion will elucidate these two figures.
First of all, the $S$-wave scattering phase shifts $\delta_0(k)$
are calculated for four specific twisting angles, $\bm{\theta} =(0,0,0)$, $(0,0,\pi)$, $(\pi,\pi,0)$, and $(\pi,\pi,\pi)$, at which Eq.~\eqref{eq:Luscher_formula_origin} is applicable. These data points are presented by circle symbols in the left panel of Fig.~\ref{fig:plot_kcot}. Within our calculated range, the quantity of $k \cot\delta_0$ is observed to increase monotonically. Consequently, the result of interpolating these four data points with the quadratic polynomial in $k^2$ is the dashed curve, and the gray shaded area represents the error obtained by the jackknife method.

Second, the $P$-wave scattering phase shift $\delta_1(k)$ is calculated through the finite-size formula~\eqref{eq:Luscher_formula_twist3d} using the energy levels measured under the twisted boundary 
conditions with a single type of the twisting angle vector, $\bm{\theta}=(\theta,\theta,\theta)$. As the $S$-wave scattering phase shift is incorporated into the right-hand side of the formula~\eqref{eq:Luscher_formula_twist3d}, the value
of $\delta_0(k)$ at a given $k^2$ is fixed by the fitting curve presented in the right panel.
The resulting $P$-wave scattering phase shifts
are presented by diamond symbols in the left panel of Fig.~\ref{fig:plot_kcot}, where the $k^2$ dependence is negligible, enabling the data to be fitted with a linear function of $k^2$ and extrapolated to the region $k^2\sim 0$. The result of interpolating four data points with
the linear function of $k^2$ is the dashed line, and the gray shaded area represents the error obtained by the jackknife method.

The calculation of the $S$-wave scattering shift near the threshold is finally possible through the utilization of the energy levels measured under the twisted boundary conditions with a different type of the twisting angle vector, ${\bm \theta}=(0,0,\theta)$, using 
the finite-size formula~\eqref{eq:Luscher_formula_twist1d}.
Just as in the one previous discussion, the value
of $\delta_1(k)$, which is incorporated into the rhs of the formula~\eqref{eq:Luscher_formula_twist1d}, is fixed by the fitting line presented in the right panel.
In the left panel of Fig.~\ref{fig:plot_kcot}, 
the resulting $S$-wave scattering phase shifts
are presented by square symbols.
These symbols are located along the dashed curve, indicating that the gap between ${\bm \theta}=(0,0,0)$ and $(0,0,\pi)$ can be filled by resolving the mixing between $S$- and $P$-wave contributions in the finite-size formula extended for the twisted boundary conditions 
as defined in Eq~\eqref{eq:Luscher_formula_twist}.
The behavior of the scattering phase shift around the threshold is of particular importance in understanding bound state formation and determining the value of the scattering length. In this study, the utilization of twisted boundary conditions plays a crucial role.

%
% FIG. 4
%

\subsection{Scattering phase shifts for the $DD^*$ and $BB^*$ systems}
In this study, we consider the two-hadron scattering for $I=0$ and $I=1$ $HH^\ast$ systems. Heavy quarks, such as charm and bottom quarks, are treated at each physical mass, while degenerate up and down quarks are simulated at three different pion masses: $m_{\pi} =$ 295, 411, and 569 MeV. Therefore, for a total of 12 combinations in terms of labels of $I$ (isospin),  $HH^\ast$ (heavy), and $m_\pi$ (light), we obtain the respective $S$- and $P$-wave scattering phase shifts using the procedure described in Sec.~\ref{sec:Strategy}.

\subsubsection{$I=0$ $S$-wave scattering}
\begin{figure*}[htb]
\centering
\includegraphics*[width=.9\textwidth]{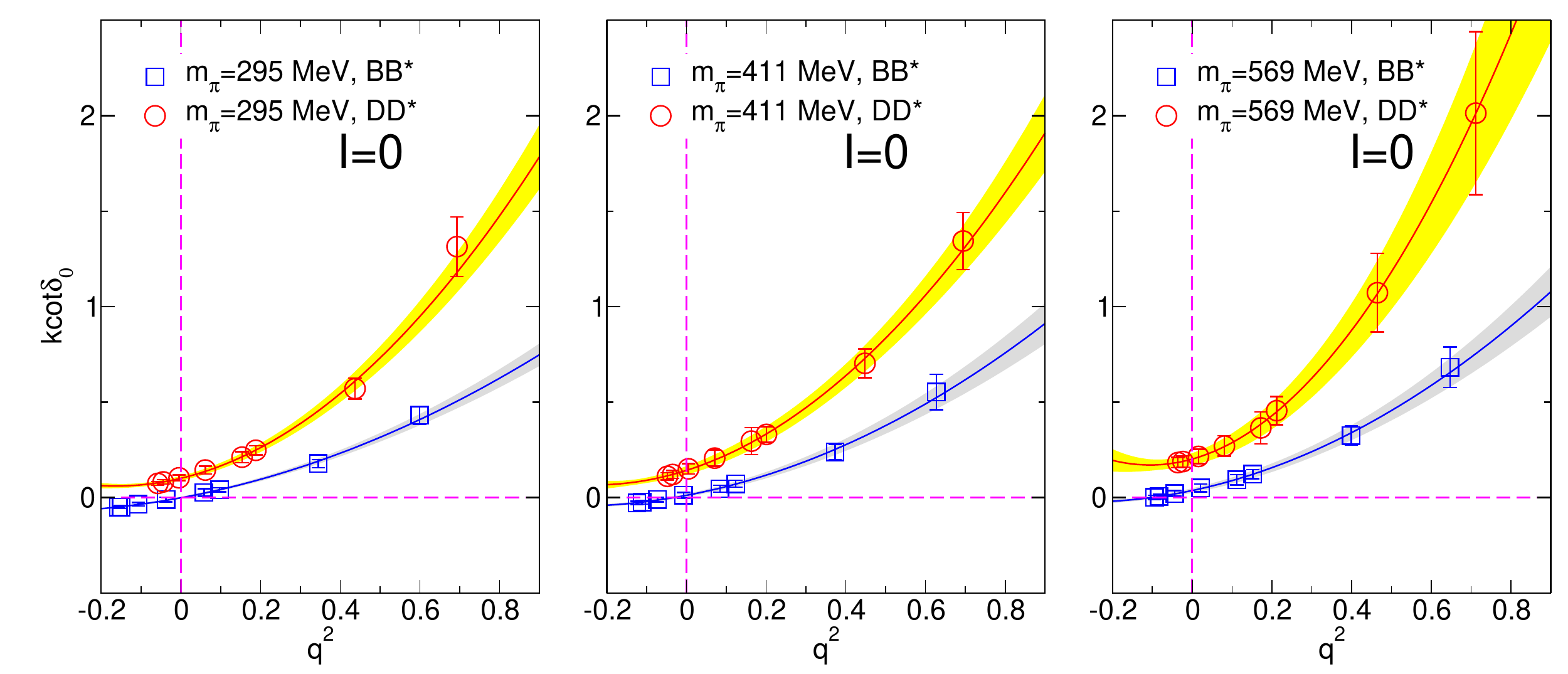} 
\includegraphics*[width=.9\textwidth]{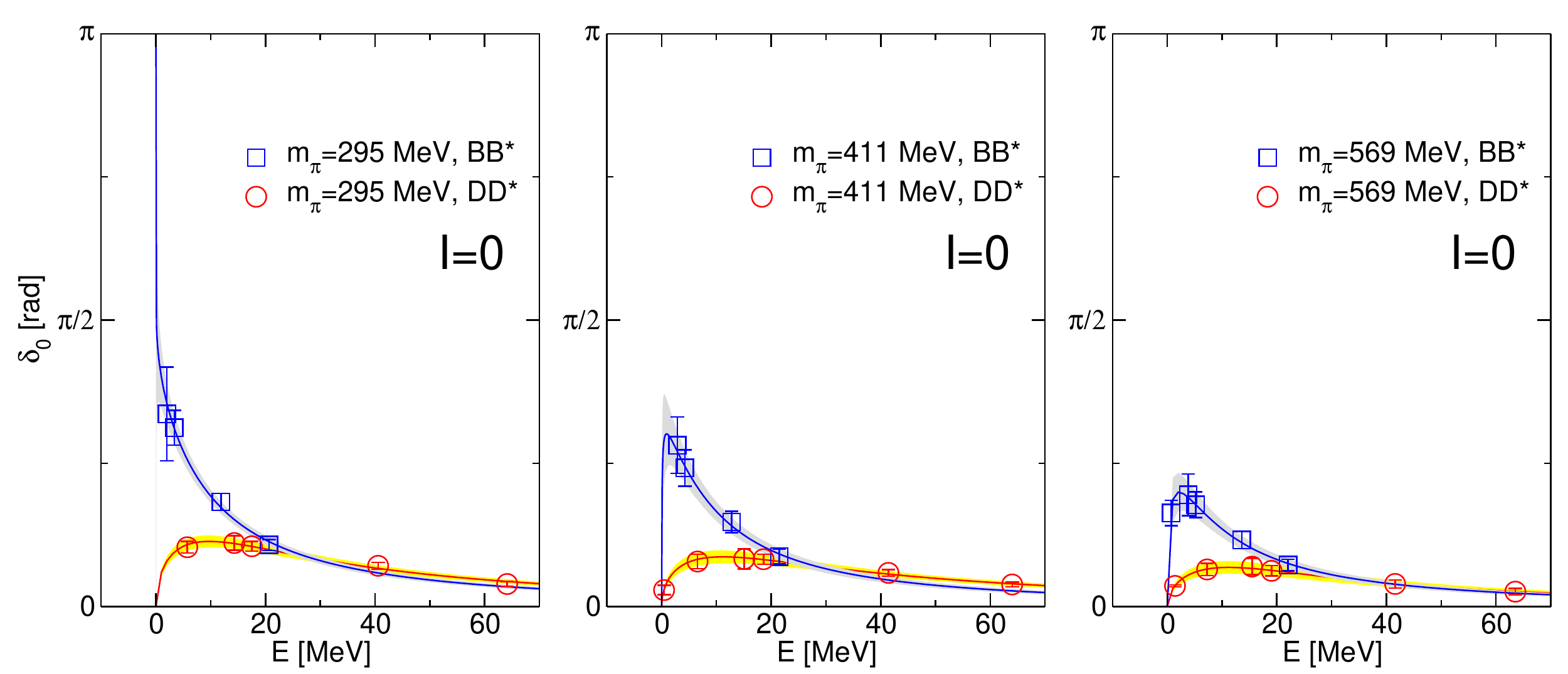} 
\caption{
$S$-wave phase shifts for the $DD^*$ and  $BB^*$ scatterings
in the $I=0$ channel:
$m_\pi=295~\mathrm{MeV}$ (left),
$m_\pi=411~\mathrm{MeV}$ (middle),
and $m_\pi=569~\mathrm{MeV}$ (right).
The upper panels illustrate the quantity $k\cot\delta_0$ as a function of $q^2$, while the phase shift $\delta_0$ is plotted as the two-hadron energy $E$, measured from the threshold, in the lower panels.}
\label{fig:plot_delta0}
\end{figure*}

Figure~\ref{fig:plot_delta0} presents the results of the $S$-wave scattering phase shifts obtained for the $I=0$ case. Three upper panels of Fig.~\ref{fig:plot_delta0}
display the results of $k\cot\delta_0$ as a function of $q^2$. The left, center and right panels show the results
obtained at $m_{\pi} =$ 295, 411, and 569 MeV, respectively. The $DD^\ast$ and $BB^\ast$ channels are denoted by circle and square symbols, respectively. 
Similarly, the results of $\delta_0$ are also plotted
as a function of the two-hadron energy $E$ measured from the respective threshold in the three lower panels of Fig.~\ref{fig:plot_delta0}. 
The scattering phase shifts are positive, reflecting the attractive interactions in the $I=0$ channel of both the $DD^\ast$ and $BB^\ast$ systems.

Suppose the interaction between the $H$ and $H^\ast$
mesons is approximately mediated by light hadrons, such as the pion and the $\rho$ meson. It is anticipated that such interaction becomes stronger as the panel moves from right to left. Indeed, the values of the phase shifts 
increase, from the right panel to the left panel,
in both the $DD^\ast$ and $BB^\ast$ cases. 
This tendency is more pronounced in the $BB^\ast$ case than the $DD^\ast$ case. Moreover, for the $BB^\ast$ case, the scattering phase increases rapidly near the threshold as $m_\pi$ decreases. Such observations made in the $BB^\ast$ channel are indicative of the bound state formation between the $B$ and $B^\ast$ mesons. 

\begin{figure}[htb]
\includegraphics[width=.98\linewidth,bb=0 0 792 612,clip]{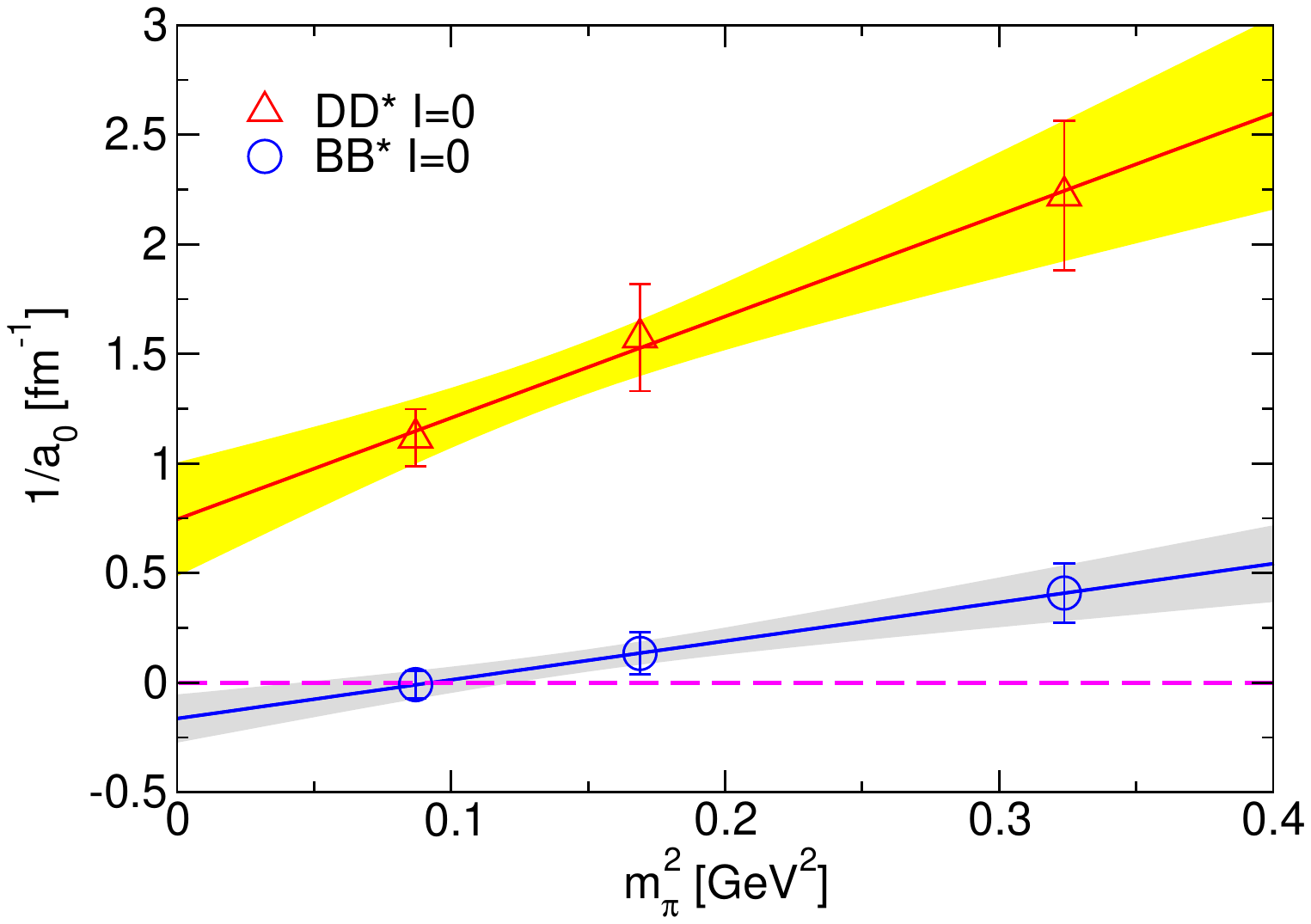} 
\caption{
The pion mass dependence of the inverse of scattering length in the $I=0$ $DD^*$ and $BB^*$ channels.
The horizontal dashed line indicates the unitary limit ($a_0=\pm \infty$).
}
\label{fig:plot_scattl_pion_mass}
\end{figure}

As demonstrated in the upper left panel, for the $BB^\ast$ case, the value of $k\cot\delta_0$ at $q^2=0$ becomes zero, thereby indicating the infinite scattering length corresponding to the unitary limit.
The inverse of the scattering length can be read off from the $y$-axis intercept by fitting all data of $k\cot \delta_0(k)$  using the form of the effective range expansion~\eqref{eq:ERE}.
Figure~\ref{fig:plot_scattl_pion_mass} shows the inverse of the scattering length as a function of $m_\pi^2$. As $m_\pi^2$ decreases, the inverse of the scattering length decreases monotonically.
Remarkably, the $I=0$ $BB^{\ast}$ system reaches the unitary limit ($a_0=\pm \infty$) at $m_\pi=$ 295 MeV and exhibits the characteristic behavior of the scattering phase shift at $E=0$ MeV, as shown in the lower left panel of Fig.~\ref{fig:plot_delta0}. The threshold phase shift begins at $\pi$, suggesting that at least one bound state forms at the lighter $m_\pi$. 

It should be remarked that the number of data points displayed in each of the upper and lower panels differs. This is simply because the corresponding data points of $k\cot \delta_0$ in the negative $q^2$ region cannot be interpreted into the scattering phase shifts. However, these data can be used to fit $k\cot \delta_0$ as a function of $q^2$, resulting in small error bands for the scattering phase shift $\delta_0$ near $E=0$ in the lower panels.

Finally, two scattering parameters---the scattering length $a_0$ and the effective range $r_0$--- for the $S$-wave scattering can be evaluated by fitting all data points of $k\cot \delta_0(k)$ using the form of the effective range expansion~\eqref{eq:ERE} up to ${\cal O}(k^6)$. The resulting values of $a_0$ and $r_0$ can be found in Table~\ref{tab:scattering_length}.
For the $DD^*$ case, our results of the scattering length
are consistent with Refs.~\cite{Padmanath:2022cvl,Collins:2024sfi,Chen:2022vpo, Ikeda:2013vwa,Lyu:2023xro}.

In the range of $m_\pi$ larger than 295 MeV used in this study, 
neither the $T_{cc}$ nor the $T_{bb}$ states were found to be deeply bound states in the attractive $I=0$ channel of the $DD^\ast$ or $BB^\ast$ systems.
However, as previously discussed, the scattering length of the $I=0$ $BB^\ast$ scattering begins to diverge at $m_\pi=295$ MeV, indicating the onset of bound state formation at the lighter $m_\pi$.
There is no necessity to behave linearly with respect to $m_\pi^2$ 
when a bound state is formed. However, the result for the $BB^\ast$ case does not show any nonlinearity at the lightest $m_\pi$, even though the scattering length reaches the unitary limit.
Therefore, we use a simple linear function of $m_\pi^2$ to extrapolate the scattering length toward the physical point.
We then obtain the $I=0$ scattering lengths for the
$DD^*$ and $BB^*$ systems at the physical pion mass $m_\pi^2=135$ MeV as
\begin{equation}
a_0^{DD^*} = 1.21(34)\mathrm{~[fm]},\quad a_0^{BB^*}=-7.6(5.8)\mathrm{~[fm]}.
\end{equation}

This simple analysis suggests that it is difficult to 
identify the experimentally observed $T_{cc}$ state 
with a hadronic $DD^\ast$ bound state, since the scattering length for the $I=0$ $DD^\ast$ scattering does not reach the unitary limit, even at physical points.
On the other hand, for the $BB^\ast$ case, at our lightest $m_\pi$, the scattering length reaches the unitary limit, and the scattering length at the physical point becomes negative.
This indicates the presence of a shallow bound state in the $BB^\ast$ system~\cite{Sasaki:2006jn}. For the shallow bound state, we can approximately estimate the binding energy $E_B$ by the scattering length $a_0$ by the following relation~\cite{Newton:1982qc}:
\begin{equation}
E_B\approx\frac{1}{2\mu a_0^2},
\end{equation}
where $\mu$ is the reduced mass of the $BB^\ast$ system.~\footnote{
For the $BB^\ast$ system, it is observed that the condition of $r_0\ll a_0$ is still satisfied at the physical point.}
Therefore, the binding energy $E_{B}$ for the hadronic $BB^\ast$ bound state is expected to be $\mathcal{O}(100)$ keV, using the above estimate of $a_0^{BB^*}$ at the physical point.

It should be noted that above mentioned finding for the $DD^\ast$ case is inconsistent with the recent HAL QCD result for the $T_{cc}$ study, in which the scattering length of the $I=0$ $DD^\ast$ scattering reaches the unitary limit near the physical point. However, our results for $a_0$ and $r_0$ 
are consistent with the results of other groups, including the older HAL QCD result, in the same $m_\pi$ region used in this study.
The key to resolving this discrepancy lies in the potential nonlinear behavior of $m_\pi^2$ in the inverse scattering lengths near the physical point. However, this study does not explore such a possibility without additional simulations with lighter values of $m_\pi$.

Instead, we address another discrepancy for the $BB^\ast$ case between our results and those reported in other studies. The results obtained from the other groups have predicted the existence of a deep bound state, designated as the $T_{bb}$ state, even at the heavier $m_\pi$. We would like to point out two technical differences between this study and others. The first is the difference in the treatment of the bottom quark. Other groups use
the NRQCD $b$ quark or the static $b$ quark to search the $bb\bar{u}\bar{d}$ tetraquark state or the hadronic $BB^\ast$ state.
On the other hand, in this study, the $DD^\ast$ and $BB^\ast$ systems are systematically compared in terms of heavy quark mass 
by using the RHQ $b$ quark with an equivalent treatment of the charm quark. As a result, we have succeeded in demonstrating that the $BB^\ast$ case is more likely to form a hadronic bound state than the $DD^\ast$ case.

Second, we only used the two-hadron interpolating operators
that represent two hadrons located at different coordinates
for both the source~\eqref{eq:twohadron_wall} and sink~\eqref{eq:twohadron_local} in order to achieve the aforementioned point. However, the other groups use multiple bases of the two-hadron interpolating operator. These bases include the ``compact'' four-quark operator, which is composed of two diquarks or two hadrons located at the same coordinates, designated as the tetraquark $T_{bb}$ state. 
The difference in interpolating operators explains why our calculation lacks such signal observed in the other
studies. As discussed in Ref.~\cite{Sasaki:2006jn}, 
spatially extended two-particle interpolating operators rarely couple to bound states, while conversely, spatially compact two-particle interpolating operators rarely couple to two-particle scattering states.  

As will be elaborated upon later in Appendix A, 
when the compact two-hadron interpolation operator is used for the sink only and the same source operator is used, the corresponding ``deeply bound state'' appears in the early time slices of the effective energy plot for both the $DD^\ast$ and $BB^\ast$ cases. The behavior of the plateau is uncertain for the $DD^\ast$ case. However, the plateau for the $BB^\ast$ case is relatively solid and appears at about 100 MeV below the threshold. 
If the observed energy shift can be regarded as a deep binding energy of ${\cal O}(100)$ MeV, it is consistent with other results on the binding energy of the $T_{bb}$ state. 
This observation suggests that the observed signature of shallow bound state formation is, in fact, the second bound state.
The existence of two bound states for the $T_{bb}$ state was recently predicted by four-body calculations of double-heavy tetraquarks in the quark model~\cite{Meng:2020knc}.

%
% FIG. 5
%

%
% FIG. 6
%
%\begin{figure*}[htb]
%\end{figure*}

\clearpage
\subsubsection{$I=0$ $P$-wave scattering}
\begin{figure*}[htb]
\centering
\includegraphics*[width=.9\textwidth]{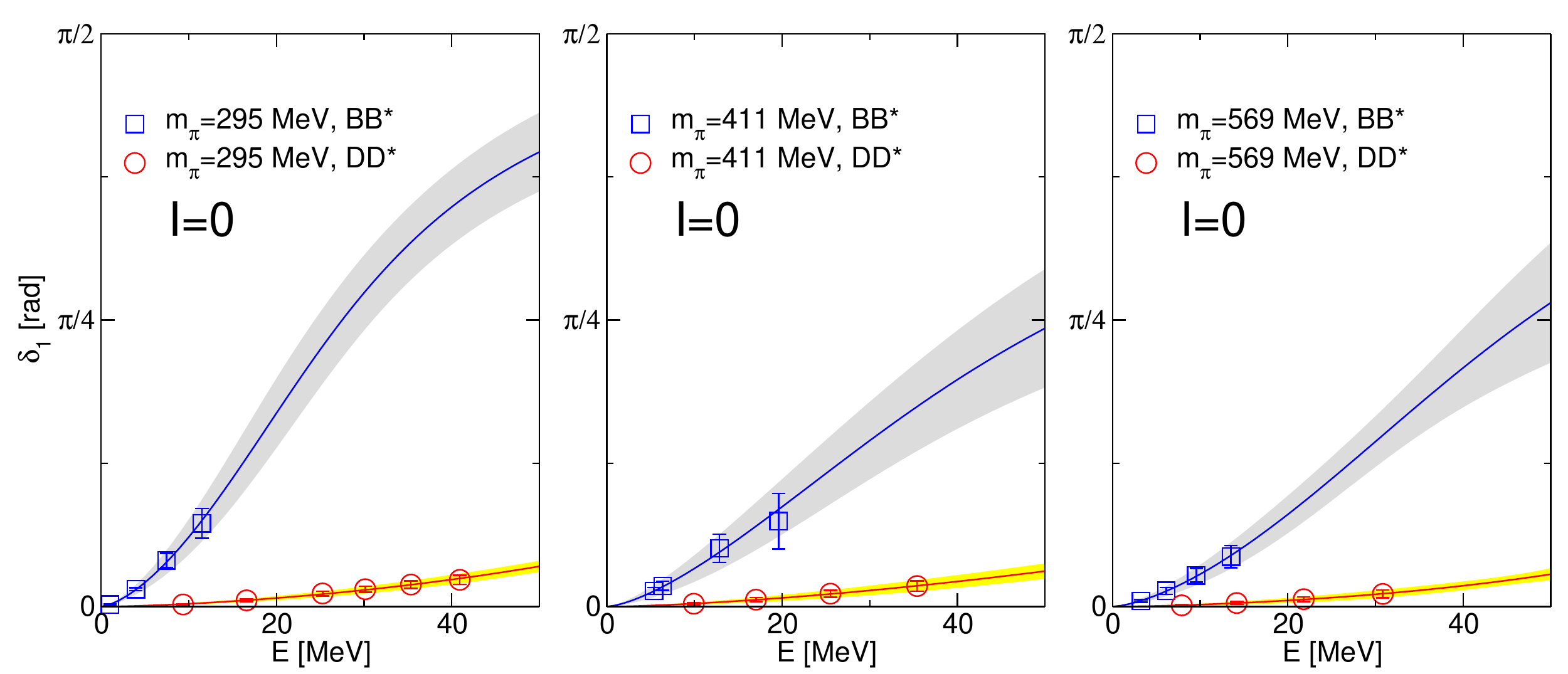} 
\caption{
$P$-wave phase shifts for the $DD^*$ and $BB^*$ scatterings
in the $I=0$ channel:
$m_\pi=295~\mathrm{MeV}$ (left),
$m_\pi=411~\mathrm{MeV}$ (middle),
and $m_\pi=569~\mathrm{MeV}$ (right).
}
\label{fig:plot_delta1}
\end{figure*}

Figure~\ref{fig:plot_delta1} shows the $P$-wave scattering phase shifts for the $I=0$ case. The results of $\delta_1$ are plotted as a function of the two-hadron energy $E$. Similar to the $S$-wave phase shifts, the $P$-wave phase shifts are
observed to be positive, reflecting the attractive interaction
in the $I=0$ channel of both the $DD^\ast$ and $BB^\ast$ systems. 

Similar to what we observed in the $S$-wave phase shift, 
the effect of varying $m_\pi$ on the $P$-wave phase shift is more pronounced in the $BB^\ast$ case compared to the $DD^\ast$ case. In particular, for the $BB^\ast$ at the smallest $m_\pi$, there appears to be a trend in which the $P$-wave phase shift near the threshold tends to increase significantly toward $\pi/2$. 
This characteristic aspect of the $P$-wave phase shift may suggest a possible precursor to the formation of a resonance state, although the mixing of higher partial waves ($l\ge 2$) that is ignored in this analysis must be considered for the $P$-wave phase shift at high energies.

The scattering parameter of the scattering volume $a_1$ for the
$P$-wave scattering is evaluated by fitting all data points
of $k^3\cot \delta_1(k)$ using the form of the effective range
expansion~\eqref{eq:ERE} up to ${\cal O}(k^4)$.
The resulting values of $a_1$ are also tabulated in Table~\ref{tab:scattering_length}.

%
% FIG. 7
%

%\clearpage
\subsubsection{$I=1$ $S$- and $P$-wave scatterings}

\begin{figure*}[htb]
\centering
\includegraphics*[width=.9\textwidth]{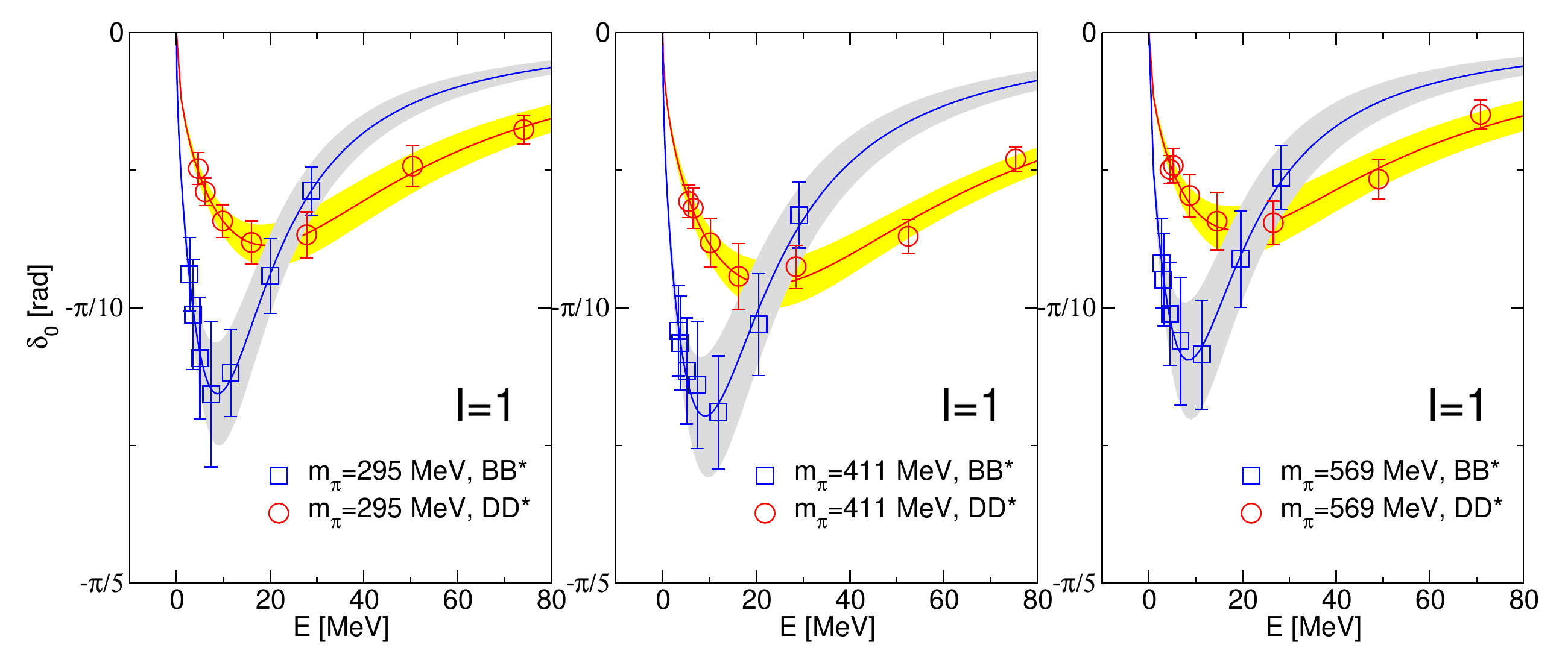} 
\caption{
$S$-wave phase shifts for the $DD^*$ and $BB^*$ scatterings
in the $I=1$ channel:
$m_\pi=295~\mathrm{MeV}$ (left),
$m_\pi=411~\mathrm{MeV}$ (middle),
and $m_\pi=569~\mathrm{MeV}$ (right).
Although $\bm{\theta}=(0,0,\theta)$ with $\theta=0.900$ is also calculated,
we exclude the data in the $I=1$ case since the condition $|q|<1/2$ is not satisfied.}
\label{fig:plot_delta0_I1}
\end{figure*}

\begin{figure*}[htb]
\centering
\includegraphics*[width=.9\textwidth]{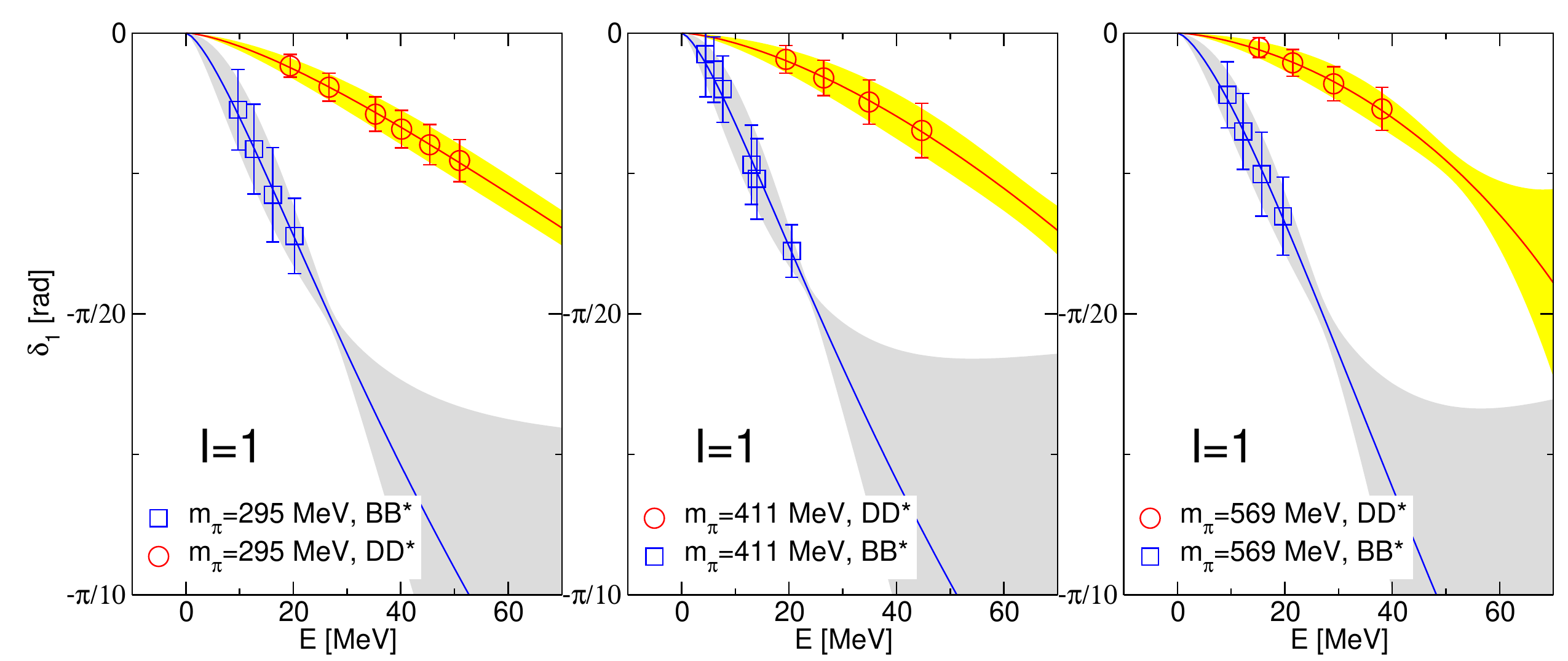} 
\caption{
$P$-wave phase shifts for the $DD^*$ and $BB^*$ scatterings
in the $I=1$ channel:
$m_\pi=295~\mathrm{MeV}$ (left),
$m_\pi=411~\mathrm{MeV}$ (middle),
and $m_\pi=569~\mathrm{MeV}$ (right).
}
\label{fig:plot_delta1_I1}
\end{figure*}

We also provide the results of the $S$- and $P$-wave scattering phase shifts for the $I=1$ case.
As demonstrated in Fig.~\ref{fig:plot_delta0_I1}, the $S$-wave phase shifts are displayed, while the $P$-wave phase shifts are presented in Fig.~\ref{fig:plot_delta1_I1}.
The negative scattering phase shifts reflect the repulsive nature of the $I=1$ channel of both the $DD^\ast$ and
$BB^\ast$ systems. The scattering parameters for both the $S$- and $P$-wave scatterings are determined in a similar manner in the $I=0$ channel, as previously outlined.
The resulting values of $a_0$, $r_0$, and $a_1$ are also included in Table~\ref{tab:scattering_length}. 

In contrast to the $I=0$ channel, the $I=1$ scattering phase shifts of the $S$ and $P$ wave are nearly independent of $m_\pi$ in both the $DD^\ast$ and $BB^\ast$ systems.
This observation makes it plausible that the strong $m_\pi$ dependence of the $S$- and $P$-wave scattering phases near threshold, as observed for the $I=0$ $BB^\ast$ case, originates from a precursor to the formation of bound and resonant states that are characteristic of attractive systems.

In addition, the absolute value of the scattering phase shifts is
much smaller than that in the $I=0$ channel. This suggests that, in the $I=1$ channel, the interaction between the $H$ and $H^\ast$ mesons is a very weak repulsive interaction.
These trends are consistent with the lattice simulations that assume the one-pion exchange model~\cite{Meng:2024kkp}.

%
% FIG. 8
%

%
% FIG. 9
%

%
% TABLE.2
%
\begin{table*}[htb]
\begin{center}
\begin{ruledtabular} 
\begin{tabular}{lccccccccc}
\hline 
&&\multicolumn{3}{c}{$I=0$} & \multicolumn{3}{c}{$I=1$} \cr
 \cline{3-5}\cline{6-8}
 System & $m_\pi$ (MeV)
 & $a_0$ (fm) & $r_0$ (fm) & $a_1$ (fm$^3$) & $a_0$ (fm) & $r_0$ (fm) & $a_1$ (fm$^3$) \cr
\hline
\multirow{3}{*}{$DD^*$}
 & 295 & 0.90(11) & 2.34(27) & 0.0209(66) & $-0.338(21)$ & $-0.07(69)$ & $-0.0218(84)$\cr
 & 411 & 0.635(98) & 3.12(54) & 0.023(10) & $-0.373(21)$ & $-0.18(62)$ & $-0.0155(89)$ \cr
 & 569 & 0.450(68) & 2.8(1.1) & 0.0137(81) & $-0.327(24)$ & $-0.6(1.4)$ & $-0.0123(80)$ \cr
 \hline
\multirow{3}{*}{$BB^*$} 
& 295 & $\pm\infty$\footnotemark[1]\footnotetext[1]{
The infinite scattering length is indicated by the observation that 
$1/a_0$ [fm$^{-1}$] $=-0.009(65)$.
} 
& 1.808(82) & 0.114(35) & $-0.434(32)$ & 1.43(66) & $-0.036(30)$ \cr
 & 411 & 7.5(5.4) & 1.85(14) & 0.070(31) & $-0.505(39)$ & 0.62(67) & $-0.040(36)$\cr
 & 569 & 2.45(79) & 2.07(22) & 0.054(21) & $-0.424(45)$ & 1.05(91) & $-0.028(20)$\cr
\hline
\end{tabular}
\caption{
Fit results of the scattering length $a_0$ and the effective range $r_0$ 
for the $S$-wave scatterings and the scattering volume $a_1$ for the $P$-wave scatterings.}
\label{tab:scattering_length}
\end{ruledtabular}
\end{center}
\end{table*}

\clearpage
%%%%%%
\section{Summary}
\label{sec:summary}
We have investigated the properties of the doubly heavy tetraquark states, $QQ \bar{u}\bar{d}$ ($Q=c$ or $b$), in lattice QCD simulations from the perspective of hadronic molecular states.
For this purpose, the $S$- and $P$-wave phase shifts for the $DD^\ast$ and $BB^\ast$ scatterings near the thresholds are studied
by the L\"uscher's finite-size method using the novel trick of twisted boundary conditions. 
Our simulations are performed using 2+1 flavor PACS-CS gauge ensembles of $m_\pi=295$, 411, and 569 $\mathrm{MeV}$ at a single
lattice spacing ($a=0.09$ fm) and volume ($La=2.9$ fm). 
In the treatment of charm and bottom quarks, the RHQ quark was commonly adopted to facilitate a systematic discussion of how the different masses of the heavy quarks can lead to distinct results.

The use of the twisted boundary conditions allows us to treat any small momentum on the lattice through the variation of the twisting angle, continuously. This method provides a significant advantage in the analysis of bound state formation, as evidenced by the behavior of the near-threshold $S$-wave scattering phase.
The introduction of the twisted boundary condition makes the mixing of even-$l$ and odd-$l$ partial waves inevitable, but this can be used to calculate the $S$- and $P$-wave scattering phase shifts simultaneously. Hence, it is also possible to search for the possible narrow resonance states near the threshold by examining the $P$-wave scattering phase shifts.

We have found that the $I=0$ channels of the $DD^\ast$ and $BB^\ast$ systems exhibit positive phase shifts of $S$- and $P$-wave scatterings, reflecting an attractive interaction. On the other hand, 
the $I=1$ channels of the $DD^\ast$ and $BB^\ast$ systems
exhibit a very weakly repulsive interaction. 
A comparison of the charm and bottom systems reveals that low-energy scattering is greater in the bottom system than in the charm system, without either $S$- or $P$-wave differences or isospin differences.

Except for the $I=0$ $BB^{*}$ system, the dependence of the light quark on low-energy scattering is not as significant as the change in the heavy quark within the $m_{\pi}$ range of 295--569 MeV.
In contrast, 
the $S$-wave phase shift of the $I=0$ $BB^\ast$ scattering 
increases rapidly near the threshold as $m_\pi$ decreases. 
Remarkably, the $I=0$ $BB^\ast$ system reaches the unitary limit at
$m_\pi=295$ MeV, causing the threshold phase shift to begin at $\pi$. This suggests that at least one bound state forms at the lighter $m_\pi$. Furthermore, the $P$-wave phase shift of the $I=0$ $BB^\ast$ scattering near the threshold tends to increase significantly toward $\pi/2$ at $m_\pi=295$ MeV.  
It may suggest a possible precursor to the
formation of a resonance state in the $I=0$ $BB^\ast$ channel. 

We also succeed in extracting model-independent information of
the low-energy $DD^\ast$ ($BB^\ast$) interaction, such as
the scattering length and the effective range for the $S$-wave and the scattering volume for the $P$ wave in both the $I=0$ and $I=1$ 
channels. As for the $S$-wave scattering length in the $I=0$ channel, our simple analysis using the linear chiral extrapolation
with respect to $m_\pi^2$ suggests the existence of a shallow bound state of the $BB^\ast$ molecule with a binding energy of $\mathcal{O}(100)$ keV. This state differs from the deeply bound state, which has an observed binding energy of $\mathcal{O}(100)$ MeV and is considered the $T_{bb}$ state, as observed in other studies. 

As demonstrated in Appendix A, the absence of the deeply bound state in our calculations is indeed the case.
It has been determined that this is due to the two-hadron interpolating operators used in this study.
These operators are specifically configured for two-hadron scattering.
In light of these implications, it is suggested that the signature of shallow bound state formation observed in the $I=0$ $BB^\ast$ channel in this study represents a second bound state.
Further verification is needed to confirm the conjecture of two bound states. This requires a more comprehensive analysis, with a large operator base that includes the compact ``diquark-antidiquark''
interpolating operators as well as the compact ``meson-meson''  interpolating operators, using the variational method.

Although the $DD^{\ast}$ state was not observed as a bound state in this study, more careful treatment of the left-hand cut is necessary for a more realistic prediction.
In future work, additional lattice simulations should be conducted to investigate the dependence of the scattering length on $m_\pi$ near the physical point.

%--- acknowledgments ------------------------------------------------  
\begin{acknowledgments}
M.~N. is supported by Graduate Program on Physics for the Universe (GP-PU) of Tohoku University.
Numerical calculations in this work were partially performed using Yukawa-21 at the Yukawa Institute Computer Facility. This work was also supported in part by Grants-in-Aids for Scientific Research from the Ministry of Education, Culture, Sports, Science and Technology (No. 22K03612) and JSPS Research Fellows (No. 24KJ0412).
\end{acknowledgments}
%----------------------------------- appendix -------------------------------------------------------  

\clearpage
\appendix

\section*{Appendix A: Operator dependence in finding a deeply bound state}
\label{appendix_A}

In the $I=0$ $bb\bar u\bar d$ channel, the existence of a deeply bound state, designated as the $T_{bb}$ state, has been reported in other calculations.~\cite{Francis:2016hui, Junnarkar:2018twb, Leskovec:2019ioa, Mohanta:2020eed, Hudspith:2023loy, Aoki:2023nzp, Alexandrou:2024iwi}.
In the search for the tetraquark $T_{bb}$ state, multiple bases of the two-hadron interpolating operator are utilized. These bases include the compact four-quark operator, which is composed of two diquarks or two hadrons located at the same coordinates. On the other hand, this study was undertaken to calculate the scattering phase shifts of both the $DD^\ast$ and $BB^\ast$ systems. To this end, the two-hadron interpolating operator composed of two hadrons located at two spatially separated coordinates was utilized (referred to as a ``spatially separated type'').

To understand that the deeply bounded $T_{bb}$ state is not visible in our study, we additionally examine the following two-hadron interpolating operator:
\begin{align}\label{eq:four_quark_compact}
\Omega^{C(I)}_{HH^{*}(i)}({\bm x};t)=\frac{1}{\sqrt2}
\left\{O^{L,u}_{H}({\bm x},t)O^{L,d}_{H^*(i)}({\bm x},t)-(-1)^IO^{L,d}_{H}({\bm x},t)O^{L,u}_{H^*(i)}({\bm x},t)\right\},
\end{align}
where the two hadrons are located at the same coordinates (referred to as a ``spatially compact type'').
Through the Fierz rearrangement of the Dirac and color indices,
above the compact meson-meson interpolating operators is related to a linear combination of the diquark-antidiquark interpolating operators.
Instead of the sink operator defined in Eq.~\eqref{eq:twohadron_local}, this operator~\eqref{eq:four_quark_compact} is used for the sink operator
to construct the four-point correlation functions for the $I=0$ two-hadron system defined as
\begin{align}
\label{eq:compact_two_hadron_correlation}
\frac{1}{3}\sum_i \frac{1}{L^3}\sum_{{\bm x}}
\langle \Omega^{C(I=0)}_{HH^{*}(i)}({\bm x},t)\Omega^{W(I=0)}_{HH^{*}(i)}(-{\bm p};t_{\mathrm{src}})^\dagger\rangle,
\end{align}
where the spatial coordinate of two hadrons at the sink are summed over simultaneously, and the total momentum of the $HH^\ast$ system is zero.
Note that the four-point correlation function becomes identically zero for the $I=1$ case. 
As demonstrated in Ref.~\cite{Sasaki:2006jn}, spatially compact two-particle interpolating operators are more strongly coupled to bound states than to two-particle scattering states.
In the following calculations, we do not perform the variational method, but simply look at the differences in the effective energy plots of the ratio correlation defined in Eq.~\eqref{eq:ratio_correlator}, with two types of the four-point correlation functions, Eq.~\eqref{eq:two_hadron_correlation} and~\eqref{eq:compact_two_hadron_correlation}.

\begin{figure*}[htb]
\centering
\includegraphics*[width=.48\textwidth,bb=0 0 792 612,clip]{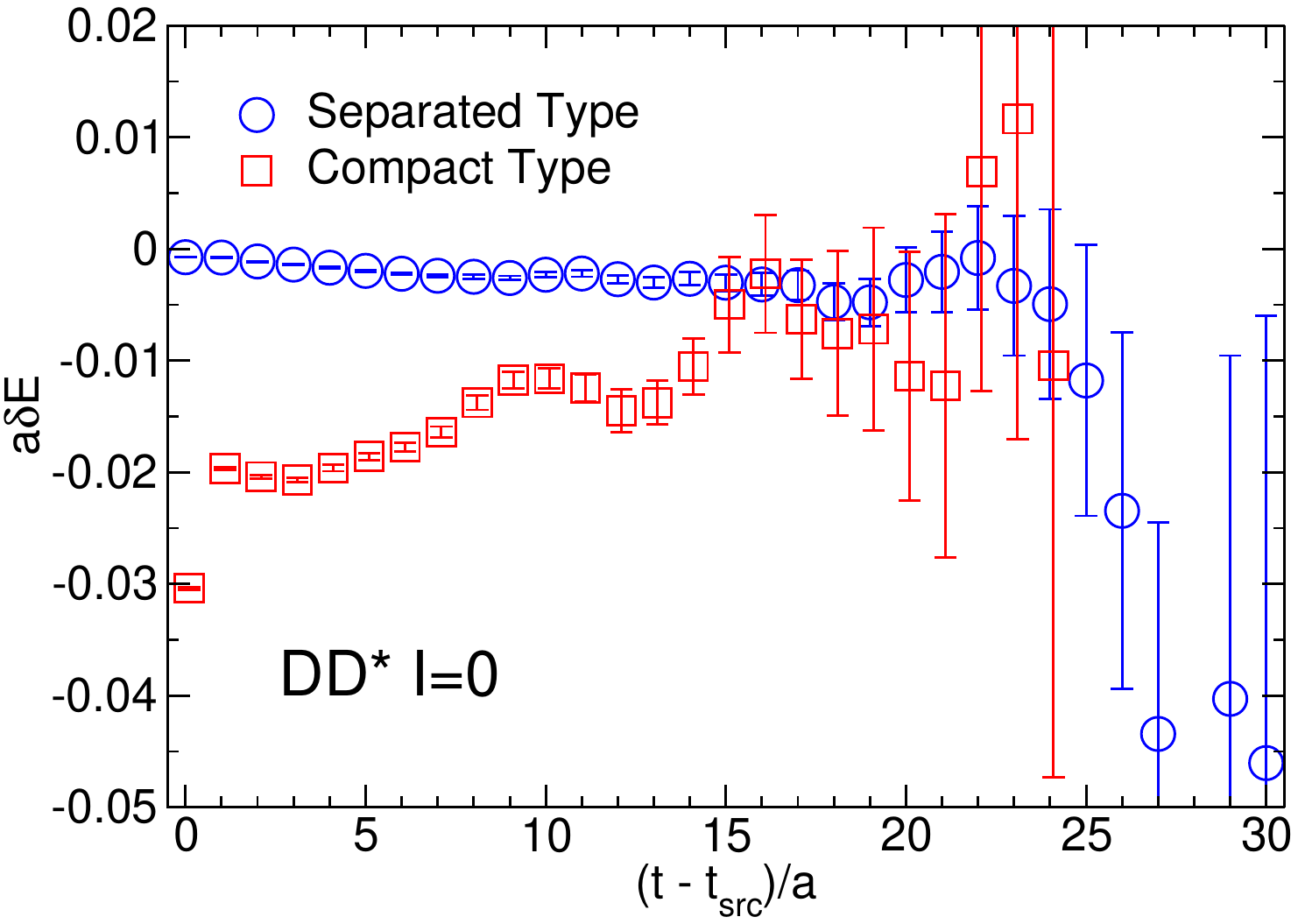} 
\includegraphics*[width=.48\textwidth,bb=0 0 792 612,clip]{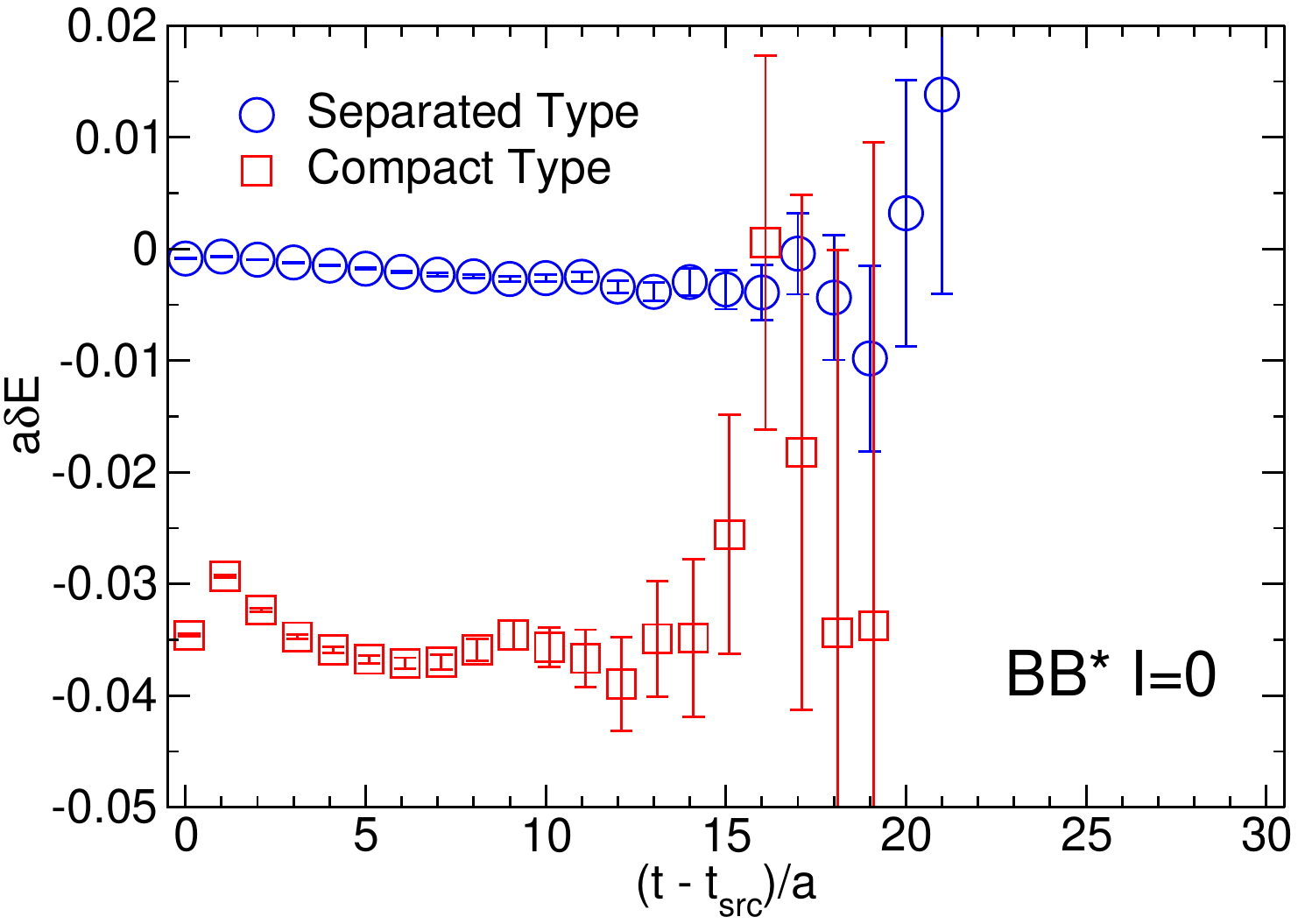}
\caption{
The effective energy shift $\delta E_{\rm eff}$ as a function of $t$ 
 for the $DD^{*}$ (left) and $BB^{*}$ (right) systems in the $I=0$ channel at $m_\pi=295$ MeV (ensemble A) with the trivial twisting angle, ${\bm \theta}=(0,0,0)$, which corresponds to the standard periodic boundary condition.
Each panel contains results obtained using two different types of the sink operator for two hadrons: a spatially separated type, as defined in Eq.~\eqref{eq:twohadron_local} (denoted by circles), and a spatially compact type, as defined in Eq.\eqref{eq:four_quark_compact} (denoted by squares).}
\label{fig:plot_DeltaE_compact}
\end{figure*}

Figure~\ref{fig:plot_DeltaE_compact} shows the effective energy $\delta E_{\rm eff}$ for the $DD^\ast$ (left) and $BB^\ast$ (right) systems in the $I=0$ channel with the twisting angle ${\bm \theta}=(0,0,0)$ at $m_\pi=295$ MeV.
The circle symbols represent the results obtained using  Eq.~\eqref{eq:two_hadron_correlation}, and the square symbols represent
the results obtained using Eq.~\eqref{eq:compact_two_hadron_correlation}. 

As shown in the left panel of Fig.~\ref{fig:plot_DeltaE_compact}, the two results from the $DD^*$ system differ in the early time slices. However, both results become consistent after the condition $(t-t_{\rm src})/a \ge 15$. This indicates that the four-point function~\eqref{eq:two_hadron_correlation} has much better overlap with the two-hadron scattering state.
On the other hand, for the $BB^{\ast}$ case the effective energy obtained using Eq.~\eqref{eq:compact_two_hadron_correlation}
forms a relatively solid plateau in the range $5\le (t-t_{\rm src})/a \le 15$,
which is well below the threshold and the results for the $BB^\ast$ scattering state obtained using Eq.~\eqref{eq:two_hadron_correlation}.
The energy shift has been observed to be of ${\cal O}(100)$ MeV, which is comparable to other results on the binding energy of $T_{bb}$ state.

If the observed energy shift can be regarded as a deep binding energy of ${\cal O}(100)$ MeV, this observation may indicate that the observed signature of shallow bound state formation discussed in the text 
is, in fact, originated from the second bound state. The existence of two bound states for the $T_{bb}$ state was recently predicted by four-body calculations of double-heavy tetraquarks in the quark model~\cite{Meng:2020knc}.

In any case, the discussion here confirms that the absence of the deeply bound state in our calculations, as outlined in the text, is attributable  to the two-hadron interpolating operator employed, which was configured specifically for two-hadron scatterings. One important lesson from this analysis is that even when deeply bound states are formed, the two-hadron interpolating operators employed in this study are not strongly coupled to those states. On the contrary, our chosen operators are well suited for calculating two-hadron scattering phase shifts. Needless to say, one should also be cautious in regard to the potential occurrence of a false plateau in multihadron spectroscopy~\cite{Iritani:2016jie}.

%
% FIG. 10
%

%--- bibliography ---------------------------------------------------  

\end{document}